\documentstyle[aps,preprint,epsf]{revtex} \draft
\tightenlines

\newcommand{\etal}{{\it et al.}}

\def\LP{\left(}		
\def\RP{\right)}	

\def\BE{\begin{equation}}
\def\EE{\end{equation}}
\def\BEA{\begin{eqnarray}}
\def\EEA{\end{eqnarray}}

\newcommand{\pbp}{\bar\psi\psi}

\begin{document}

\title{
The QCD spectrum with three quark flavors }

\author{ Claude~Bernard }
\address{
Department of Physics, Washington University, St.~Louis, MO 63130, USA
}
\author{ Tom Burch, Kostas Orginos\footnote{Present address: RIKEN-BNL Research Center,
Brookhaven National Laboratory, Upton, NY 11973-5000} and Doug~Toussaint }
\address{
Department of Physics, University of Arizona, Tucson, AZ 85721, USA
}
\author{ Thomas~A.~DeGrand }
\address{
Physics Department, University of Colorado, Boulder, CO 80309, USA
}
\author{ Carleton~DeTar }
\address{
Physics Department, University of Utah, Salt Lake City, UT 84112, USA
}
\author{ Saumen Datta and Steven~Gottlieb}
\address{
Department of Physics, Indiana University, Bloomington, IN 47405, USA
}
\author{ Urs~M.~Heller }
\address{
CSIT, Florida State University, Tallahassee, FL 32306-4120, USA
}
\author{ Robert~Sugar }
\address{
Department of Physics, University of California, Santa Barbara, CA 93106, USA
}

\date{\today}

\maketitle

\newpage
\begin{abstract}\noindent
We present results from a lattice hadron spectrum calculation using
three flavors of dynamical quarks ---  two light and one strange, and
quenched simulations for comparison.  These simulations were done
using a one-loop Symanzik improved gauge action and an improved Kogut-Susskind
quark action.  The lattice spacings, and hence also the physical volumes, were
tuned to be the same in all the runs to better expose differences due to flavor
number.  Lattice spacings were tuned using the static quark potential, so 
as a byproduct we
obtain updated results for the effect of sea quarks on the static quark potential.
We find indications that the full QCD meson spectrum is in better agreement with
experiment than the quenched spectrum.  For the $0^{++}$ ($a_0$) meson we
see a coupling to two pseudoscalar mesons, or a meson decay on the lattice.
\end{abstract}

\pacs{11.15Ha,12.38.Gc}

%
%
%
%
%
%
%
%

\narrowtext
\section{Introduction}

Computation of the properties of hadrons beginning from the QCD Lagrangian
is a major goal of lattice gauge theory, and steady progress has been made.
The computational burden of including dynamical quarks is a major obstacle
in the use of lattice QCD to compute hadronic properties.  As a result, many
quantities are much better determined in the quenched approximation than in full
QCD.  One way of studying the effects of dynamical quarks is to calculate
in quenched and full QCD, using the same valence quark action in both cases, 
and matching the lattice spacings and physical sizes of the lattices, so
that any differences that are found can convincingly be ascribed to the
dynamical quarks.  Here we present a calculation of the hadron spectrum
at a lattice spacing of about $a=0.13$ fm, using quenched and full QCD lattices
at the same lattice spacing.
The lattice spacing was tuned by making short runs on smaller
lattices, adjusting the parameters to match the lattice spacing
of an initial quenched run at $10/g^2=8.0$.

We use an improved Kogut-Susskind quark action which removes tree level
order $a^2$ lattice artifacts\cite{IMP_ACTION}.
The gauge action is a one-loop Symanzik improved action\cite{SYM_GAUGE}.
This action has been shown to reduce flavor symmetry breaking and to
improve rotational symmetry of the hadron spectrum, and
to give improved scaling of hadron masses as a function of lattice
spacing\cite{IMP_SCALING}.

Another important improvement of these calculations over previous generations
is that we use three flavors of dynamical quarks.  For quark masses larger
than the strange quark mass we use three degenerate flavors, and for light
quark masses less than $m_s$ we use two light flavors, keeping the third
quark mass at about the strange quark mass.
We have also done one two-flavor simulation on a matched lattice to check
for effects of the dynamical strange quark.
For the runs with $2+1$ dynamical flavors, we computed hadron spectra
using valence quark masses equal to the sea quark masses.  In the
quenched run we computed hadron masses with
valence quarks with the same masses, and nondegenerate propagators
using a strange quark mass of $am_s=0.05$.  Finally, in the two dynamical
flavor run we computed hadron propagators using light quark masses
equal to the dynamical mass, $am_{u,d}=0.02$ and a non-dynamical strange
quark with $am_s=0.05$.

Two issues that we do not completely address are extrapolation to zero lattice
spacing and extrapolation to the physical light quark mass.
Using matched quenched and full QCD
lattices allows us to draw conclusions about the effects of dynamical
quarks without explicit continuum extrapolation.  Although it is
in principle possible that the discretization errors in quenched
and full QCD are very different, we expect that the differences
in lattice spacing dependence are in fact comparable to the differences
in physical quantities themselves.  Since corrections
to scaling are in any case quite small with our current improved 
action\cite{IMP_ACTION,IMP_SCALING} (see also below), we confidently
expect conclusions drawn 
at fixed lattice spacing to survive in the continuum limit.
We are beginning a series of simulations at a smaller lattice
spacing which will eventually allow us to make the continuum extrapolation.
A few preliminary quenched points from these finer lattice runs are included
here to give an idea of the size of these effects.
A complete chiral extrapolation will be more difficult.  In this work, we
attempt an explicit chiral extrapolation only for the shape of the static
quark potential, and show other quantities as functions of the quark mass.
Some quantities such as ``$J$''\cite{UKQCD_J} are only minimally sensitive
to chiral extrapolation, and such quantites provide immediately useful tests
of dynamical quark effects.

In addition to presenting hadron spectra, we update our computation of the
static quark potential.  For this quantity we have sufficiently accurate data
that we can hazard an extrapolation to zero quark mass to produce numbers
that can be compared with phenomenological potential models.  Because the static
potential is determined very accurately, it clearly shows differences between
quenched and full QCD.   In fact, one can even see the differences between two
and three dynamical flavors, and a ``kink'' in the  plots where we change from
three degenerate flavors to two light and one heavy indicates a noticeable
difference between two light plus one heavy and three light dynamical flavors.

In the meson sector we find differences between full and quenched QCD.  A nice
way of exposing these differences is the ratio $J$ proposed by Lacock and Michael\cite{UKQCD_J}.
We find that this quantity is in better agreement
with experiment in full QCD than in the quenched approximation,
as predicted in \cite{UKQCD_J}.
This is consistent with results of the CP-PACS\cite{CPPACS_EFFECTS} and
JLQCD\cite{JLQCD_EFFECTS} collaborations, who also conclude that inclusion of
two flavors of dynamical quarks improves agreement of the lattice spectrum with
the real world.

In the isovector $J^{PC}=0^{++}$ ($a_0$) channel we find a large difference between
quenched and three-flavor results.  We interpret the three flavor results as an
avoided level crossing between the $0^{++}$ meson and a two pseudoscalar state.
In other words, we see the $a_0$ decay to two mesons.

We include tables of the mass fits we have chosen, so the reader can compute 
his or her own favorite mass ratios.

\section{Simulation parameters}

For our two and three flavor simulations we used the standard hybrid-molecular
dynamics ``R algorithm'', \cite{R_ALGORITHM}
with one pseudofermion field for runs with degenerate quarks, and two
pseudofermion fields for runs with different strange and up and down
quark masses.
In all cases, we used trajectories with unit length in the simulation time.
Basic parameters of these runs are summarized in Table \ref{RUN_TABLE}.

\begin{figure}[tbp]
\epsfxsize=3.0in \epsfysize=3.0in
\rule{0.1in}{0.0in}\hspace{1.0in}\epsfbox[0 0 4096 4096]{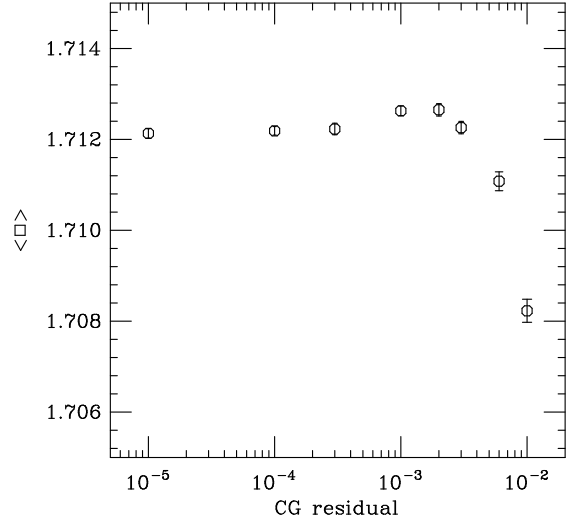}
\caption{
\label{PLQ_RES_FIG}
The effect of the conjugate gradient error used in the updating on the plaquette
in a three flavor run with quark masses $0.4 m_s$ and $m_s$.
}
\end{figure}

\begin{figure}[tbp]
\epsfxsize=3.0in \epsfysize=3.0in
\rule{0.1in}{0.0in}\hspace{1.0in}\epsfbox[0 0 4096 4096]{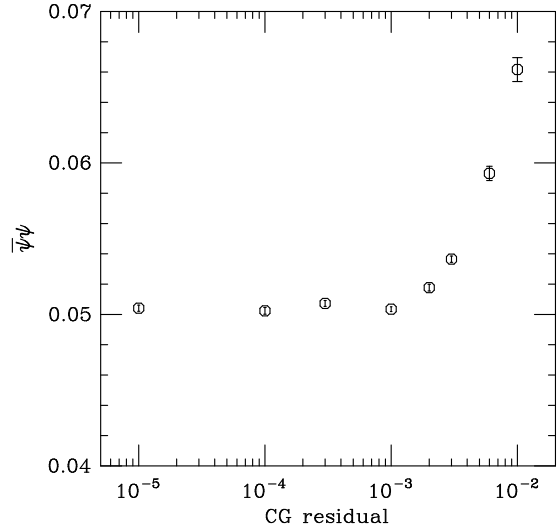}
\caption{
\label{PBP_RES_FIG}
The effect of the conjugate gradient error used in the updating on $\pbp$
in the same three flavor run.
}
\end{figure}

\begin{figure}[tbp]
\epsfxsize=3.0in \epsfysize=3.0in
\rule{0.1in}{0.0in}\hspace{1.0in}\epsfbox[0 0 4096 4096]{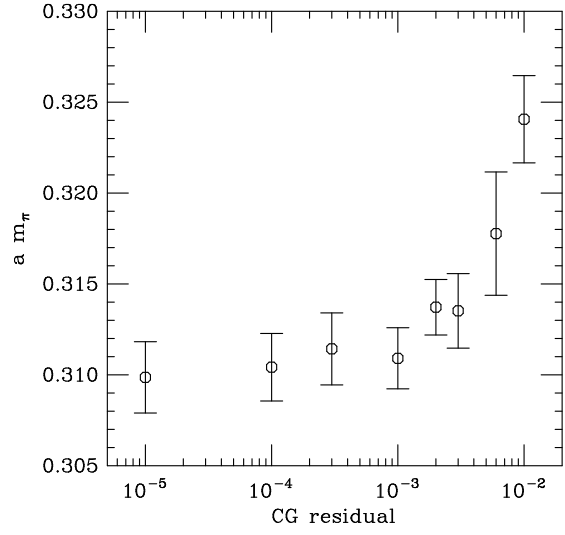}
\caption{
\label{MPI_RES_FIG}
The effect of the conjugate gradient error used in the updating on 
the Goldstone pion mass in the same three flavor run.
}
\end{figure}

Two sources of systematic error in this method are the accuracy of the 
approximate
sparse matrix solution required at each time step in the integration
of the molecular dynamics equations, and the effect of the
nonzero time step used in the integration.  We have investigated these 
effects in
the 2+1 flavor simulations at $10/g^2=6.8$ with two flavors of quarks at mass
$am_{u,d}=0.02$ and one flavor with $am_s=0.05$ on a $16^3\times 48$ lattice.
(These masses are approximately 0.4 times
the physical strange quark mass and the physical strange quark mass, 
respectively).  Figures \ref{PLQ_RES_FIG}
and \ref{PBP_RES_FIG} show the plaquette and $\pbp$ as a function of the 
conjugate gradient residual used in the updating.
Finally, to see how this effect propagates into the hadron masses, we show the 
Goldstone pion mass in these same runs in Fig. \ref{MPI_RES_FIG}.
Since the effect of this error is poorly understood,
we adopted a conservative choice of $1 \times 10^{-4}$ or $5 \times 10^{-5}$ 
in most
of our runs, and used $2 \times 10^{-5}$ for the heaviest quark mass 
($am_q=0.40$) where convergence was very fast.

\begin{figure}[tbp]
\epsfxsize=7.0in
\epsfysize=7.0in
\rule{0.1in}{0.0in}\hspace{-1.25in}\epsfbox[0 0 4096 4096]{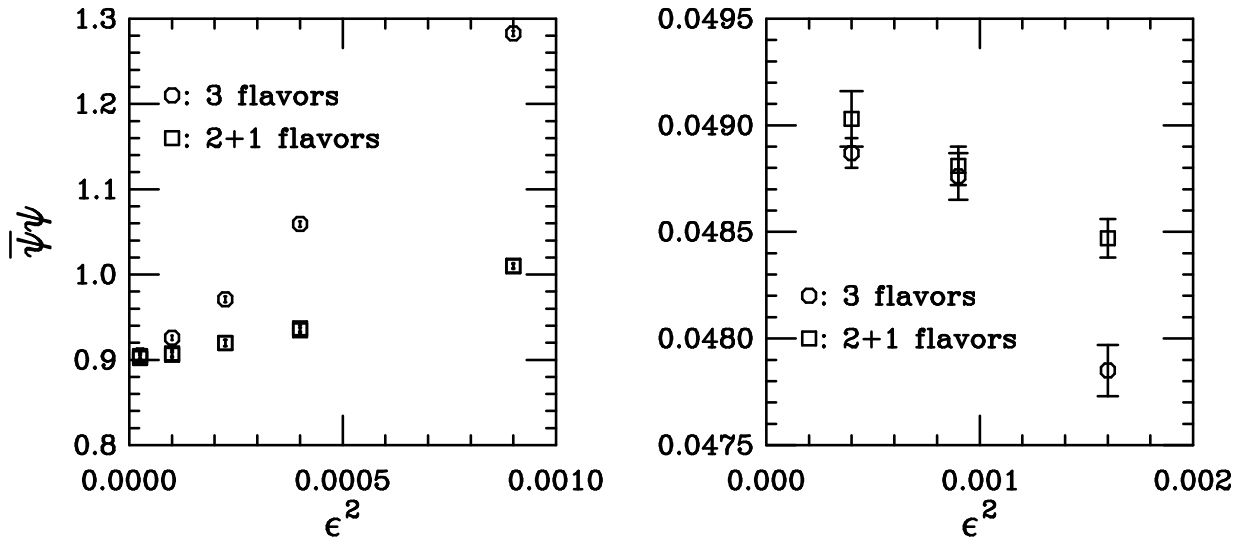}
\vspace{-3.0in}
\caption{
$\langle\bar\psi\psi\rangle$ versus the squared step size.  The left hand
panel is an algorithm test done on a $12^4$ lattice using the one-plaquette
gauge action and conventional quark action at $6/g^2=5.10$ with three quark
flavors with mass $am_q=0.02$.  The octagons use one pseudofermion field with
a factor of $3/4$ in the force term, appropriate for three flavors, while the
squares use the $2+1$ flavor code, with separate fermion force terms for
one and two flavors, but with the same mass for both terms.
The right hand panel shows $\langle\bar\psi\psi\rangle$ using improved
gauge and quark actions at $a \approx 0.14$ fm. ($10/g^2=6.80$ and $am_{u,d}=am_s=0.05$).
\label{PBP_STEP_FIG}
}
\end{figure}

The effect of integration step size is better understood.  
We verified the expected quadratic
dependence of the error on the step size in our three flavor code. 
The first panel in Fig.~\ref{PBP_STEP_FIG} shows the result of one
such test, where we ran on a quite coarse lattice using the conventional
action, comparing the old code with three degenerate flavors to the
two-plus-one flavor code with $m_{u,d}=m_s=0.02/a$.
(The ``old code'' uses a single pseudofermion vector, with a weight
of $3/4$ in the fermion force, while the ``two-plus-one flavor'' code
uses two pseudofermion vectors, one with a weight of $2/4$ and the other
with weight $1/4$.  In this test, both pseudofermion vectors had the
same mass.  In each case, the multiplication of the Gaussian random
vector by $M^\dagger$ to produce the pseudofermion vector was done
at the appropriate point in the time step to make the error in a single
time step order $\epsilon^3$, making the accumulated error over a
trajectory order $\epsilon^2$.\cite{R_ALGORITHM}.)
It can be seen that
both sets of points extrapolate to the same limit, although the size of
the effect is quite different.  The right hand panel of Fig.~\ref{PBP_STEP_FIG} shows
a similar plot from one of our pre-production tuning runs with the improved
gauge and quark action, at $10/g^2=6.80$ and $am_q=0.05$, on a $16^3\times 48$
lattice.  This is close to the value of $10/g^2$ used in the production run at
this quark mass.  Note that the finite step size corrections are quite
sensitive to the action being used -- even the sign of the effect differs
between the two actions.  
Based on these
tests and previous experience, we used a step size of
no more than two-thirds of the light quark mass or 0.03, whichever was smaller, 
in our production runs.
In the production run at $am_q=0.05$ we used a step size of $\epsilon=0.02$.  From
the slope in Fig.~\ref{PBP_STEP_FIG} we can infer that this caused a fractional
error in $\langle\bar\psi\psi\rangle$ of about 0.004, and a similar analysis
for the plaquette suggests a fractional error of about 0.0006.  We also looked
at pion masses and some of the Wilson loops involved in the computation of
the static quark potential, and we were unable to resolve statistically
significant effects on these quantities in our tuning runs.

In our production runs with dynamical quarks we measured the potential
and the spectrum at intervals of six simulation time units, and
archived these lattices.  The autocorrelation of the plaquette at
six time units, or successive measurements, was generally about 0.1.
We investigated the effect of autocorrelations on the potential and
spectrum by blocking together different numbers of measurements
before doing the fitting.  For the potential measurements we
chose to block five measurments (30 time units).  For the hadron
spectrum most of the particles showed no systematic effects of
blocking measurements together.  The exception was the pseudoscalar
mesons, where blocking gave significantly better confidence levels
and larger error estimates.  For the pseudoscalars we chose to block
together four measurements, or 24 time units.

The physical size of our lattices is $La = 20\,a \approx 2.6$ fm, which
is similar to or larger than other recent full QCD simulations.
Basic parameters of several of these calculations are summarized
in Ref.~\cite{AOKI_REVIEW}.

\widetext
\narrowtext
\widetext
\narrowtext
\widetext
\narrowtext

\widetext
\narrowtext

\section{Length scales from the static potential}

The static quark potential is often used to define the length scale
in lattice simulations.  Advantages of using the potential include ease
and accuracy of its computation, and its lack of dependence on the valence quark mass.
In comparing quenched and full simulations, subtleties arise because
the potential does depend on the masses of the sea quarks.  In Ref.~\cite{MILC_POTENTIAL}
we demonstrated the effects of sea quarks on the potential using this
improved action.  Because these effects are important in our analysis of
the hadron spectrum, we update and extend these results here.  Our methods
for computing the potential and our reasons for using $r_1$, a variant of
the conventional $r_0$ \cite{SOMMER_R0}, are described in \cite{MILC_POTENTIAL}.
$r_0$ is conventionally defined by $r_0^2 F(r_0)=1.65$, and $r_1$
by $r_1^2 F(r_1)=1.00$.

The fitting form used here is slightly more complicated than the form used in
Ref.~\cite{MILC_POTENTIAL}, with an extra term to take into account lattice
artifacts at the shortest distances.
Following a procedure used in Ref.~\cite{UKQCD_POTFORM},
\BE V(\vec r) = C + \sigma r - \alpha/r + \lambda \LP V_{free}(\vec r) - 1/r \RP \ \ . \EE
The last term, used for $r < 2.5$, approximately compensates for remaining
lattice artifacts.  Here $V_{free}(\vec r)$ is the potential calculated in
free field theory, using the improved gauge action.  Adding this term to the
fits significantly improves the goodness of fit and makes the fit parameters less
sensitive to the choice of distance range.  For the $a\approx 0.13$ fm runs we
typically find $\lambda \approx 0.3 - 0.4$.

\begin{figure}[tbp]
\epsfxsize=4.0in
\epsfysize=4.0in
\rule{0.1in}{0.0in}\hspace{1.0in}\epsfbox[0 0 4096 4096]{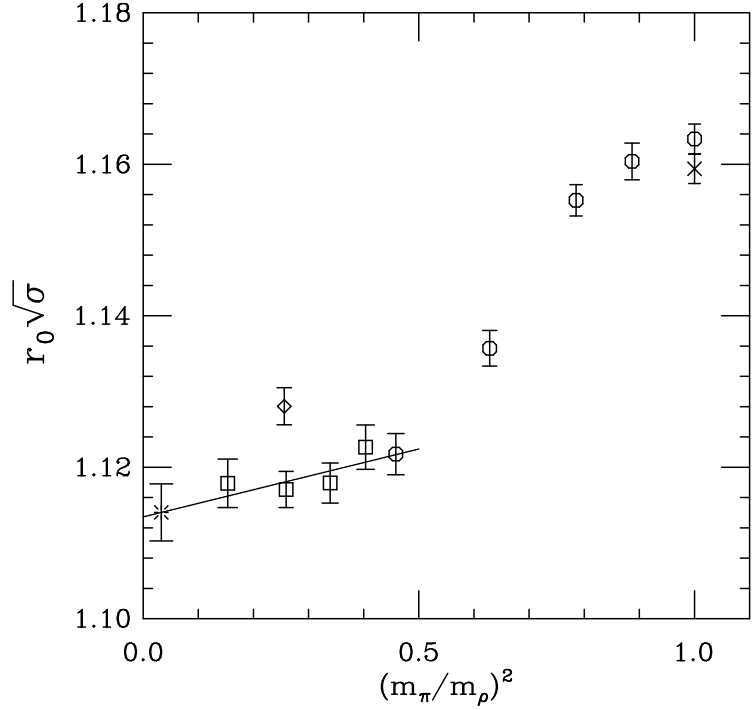}
\caption{
\label{R0_SIGMA_FIG}
Effects of dynamical quarks on the shape of the potential.  Here we plot
$r_0 \sqrt{\sigma}$ as a function of the quark mass.  The two quenched points
are at the far right, with the octagon coming from the $10/g^2=8.0$ run and
the cross from the $10/g^2=8.4$ run, which has a lattice spacing
of about 0.09 fm.  The remaining octagons are
full QCD runs with three degenerate flavors, and the squares are full
QCD runs with two light flavors and one heavy.
The diamond is the two flavor run, and the burst at the left is a linear
extrapolation of the $2+1$ results to the physical value of $(m_\pi/m_\rho)^2$.
}
\end{figure}

\begin{figure}[tbp]
\epsfxsize=4.0in
\epsfysize=4.0in
\rule{0.1in}{0.0in}\hspace{1.0in}\epsfbox[0 0 4096 4096]{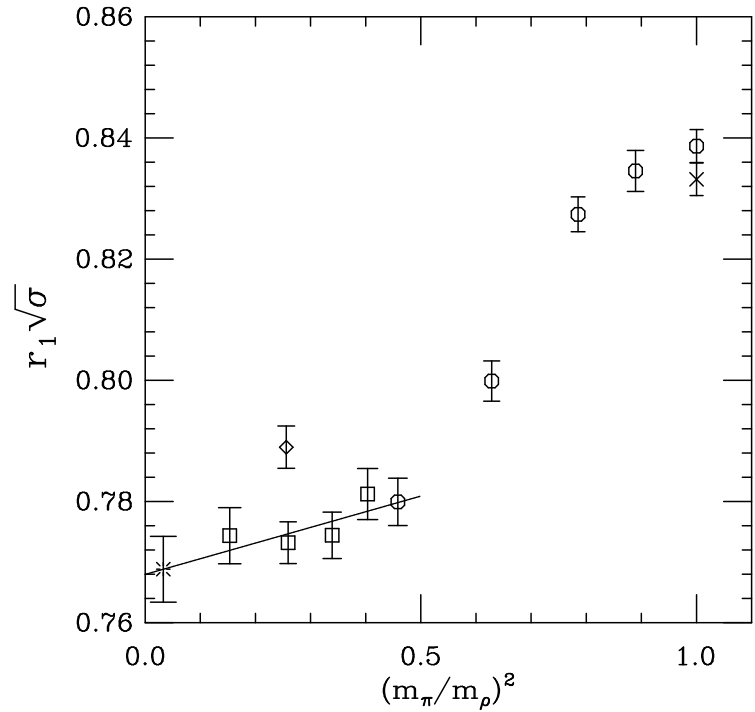}
\caption{
\label{R1_SIGMA_FIG}
The same as Fig.~\protect\ref{R0_SIGMA_FIG}, except we plot
$r_1 \sqrt{\sigma}$.  Physically, the difference is that this quantity
is sensitive to shorter distances than $r_0 \sqrt{\sigma}$.
}
\end{figure}

In Figs.~\ref{R0_SIGMA_FIG} and \ref{R1_SIGMA_FIG} we show the dimensionless
quantities $r_0\sqrt{\sigma}$ and $r_1\sqrt{\sigma}$ respectively as functions
of the quark mass, represented by $(m_\pi/m_\rho)^2$.  This places the quenched
approximation at $(m_\pi/m_\rho)^2=1$, and the chiral limit at the left side of
the graph.  In these plots the octagons are runs with three degenerate sea quarks,
except for the rightmost point which is the quenched limit. Squares
are runs with $am_s=0.05$, its approximate physical value, and 
$am_{u,d} < 0.05$.  The isolated diamond is our two flavor
run.  Finally, the cross at $(m_\pi/m_\rho)^2=1$ is the finer lattice 
quenched run.
From the two quenched points we see that remaining lattice artifacts are small
compared with the effects of the sea quarks.
In particular, the central values for $r_0\sqrt{\sigma}$ and $r_1\sqrt{\sigma}$
changed by less than 1\% when the lattice spacing was reduced from
0.13 fm to 0.09 fm, a change of about 35\%.
The kink in the plots at 
$(m_\pi/m_\rho)^2 \approx 0.46$ ($am_{u,d}=0.05$)  shows the transition 
between three degenerate
flavors and ``$2+1$'' flavors.  We can clearly see the distinction between two and
three flavors, as well as the effect of using two light and one heavy flavor rather
than three degenerate flavors (the ``kink' at $(m_\pi/m_\rho)^2 \approx 0.46$).

If we extrapolate $r_0 \sqrt{\sigma}$ to the physical quark mass, as shown in
Fig.~\ref{R0_SIGMA_FIG}, we find $r_0 \sqrt{\sigma} = 1.114(4)$ (statistical errors
only), a number which can be compared with phenomenological potential models.
The two quenched points give some idea of the possible systematic error.
Since the squared lattice spacing in the finer lattice is about one half that
of the coarser, we might expect a shift of about twice the separation of these
points in the continuum limit.  Since the error bars on these points overlap,
we don't know this systematic error well enough to justify such an extrapolation
at this point.  We expect that the effect of this systematic error is mostly
an overall shift of the graphs, but the next generation of simulations should
clarify this.

\begin{figure}[tbp]
\epsfxsize=4.0in
\epsfysize=4.0in
\rule{0.1in}{0.0in}\hspace{1.0in}\epsfbox[0 0 4096 4096]{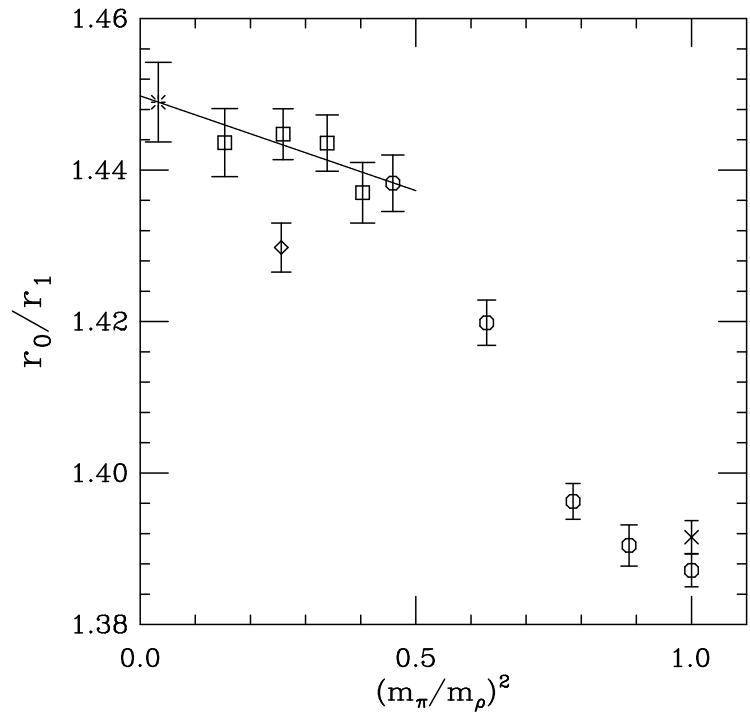}
\caption{
\label{R0_OVER_R1_FIG}
The ratio $r_0/r_1$, and a chiral extrapolation.
The symbols are the same as in Fig.~\protect\ref{R0_SIGMA_FIG}.
}
\end{figure}

While using $r_1$ to define the length scale has the advantage that it can be
done more accurately, $r_0$ has the advantage that it has been related to
phenomenological potential models, which consistently place it around 
0.5 fm\cite{SOMMER_R0}.
Therefore, to estimate $r_1$ in physical units, we plot $r_0/r_1$ in Fig.~\ref{R0_OVER_R1_FIG}.
Extrapolating linearly in $(m_\pi/m_\rho)^2$ to the physical value gives
$r_0/r_1 = 1.449(5)$ (statistical error only), or with $r_0=0.5$ fm, $r_1 = 0.35$ fm.

\widetext
\narrowtext

\widetext
\narrowtext

\widetext
\narrowtext

\section{Procedures for determining masses}

All of our hadron propagators used wall sources and local sink operators.
Several different wall sources were used.   For the ``pointlike'' hadrons,
for which all the quarks can be on a single corner of the hypercube, a ``corner
wall'' source gave the best results.  This source is simply a 1 on each
$(0,0,0)$ corner of the $2^3$ cubes on the chosen time slice.
However, to isolate the decuplet baryons a non-pointlike source is 
essential, and we used ``even'' and ``odd'' wall sources, where  $1$
or $(-1)^{x+y+z}$ is placed on each site, respectively.  This
set of sources was developed in Ref.~\cite{LANL_SPECTRUM}.
For the nonlocal pseudoscalar and vector mesons we used two wall sources
made from empirically determined linear combinations of the nonlocal pion
operators.   Finally, for the nonzero momentum mesons we used a
quark source with $1$ on each site, and an antiquark source with $e^{i \vec k \cdot\vec x}$
on each site.
All of the configurations were gauge fixed to the Coulomb gauge before 
computing the propagators.

In most cases we computed propagators from four source times evenly
spread through the lattice (only one source slice was ``turned on'' at a 
time).  For the corner source we used eight source time slices for the
light quark particles with $am_q \le 0.04$, and we used eight source
time slices for the ``even'' and ``odd'' source baryon propagators.
(The $\Delta$ propagator is very noisy, and propagators computed from
source times separated by $8 \times 0.13$ fm were basically independent.)

For Kogut-Susskind quarks the meson propagators have the generic form
\BE
{\cal H}(t) = \sum_i A_i \LP e^{-m_i t} + e^{-m_i(N_t-t)} \RP
+
\sum_i A_i^\prime (-1)^t \LP e^{-m_i^\prime t} + e^{-m_i^\prime(N_t-t)} \RP
\ \ \ .\EE
Here the oscillating terms correspond to particles with opposite parity
from the ordinary exponential terms.
Baryon propagators are similar, but have antiperiodic boundary conditions
and the ``backwards'' terms include an extra factor of $(-1)^t$.
In most cases only one mass with 
each parity is included in the fits, but for half of the pseudoscalar 
meson operators the opposite parity terms are not present, and for the 
P-wave mesons we found it necessary to keep two simple exponentials. 
The quantum numbers for the various
operators are tabulated in Ref.~\cite{KS_OPERATORS}.

Hadron masses were determined from fits to propagators, using the full
covariance matrix to estimate errors.  The maximum time distance used
in the fits was chosen to include points with fractional error less
than 0.3.   Because of the oscillating components in the staggered quark
propagators, it sometimes happens that the fractional error exceeds the
threshold at one distance but is smaller at larger distances, so the complete
criterion for the maximum distance included is the largest distance such
that the fractional error on each of the next two points exceeds 0.3.
Since the points at largest distance contribute little information, the
exact large distance cutoff is not critical.  To choose the minimum distance
included, we first went through the fits and chose a minimum distance for
each hadron in each dataset, choosing a distance where the confidence level
was reasonable and where the mass appeared to reach a plateau.
As expected, some propagators had larger(smaller) fluctuations than other
similar propagators, resulting in the choice of a larger(smaller) minimum
distance.
For the particles of greatest interest, to reduce this effect we then
``smoothed'' these minimum distances, requiring that the minimum distances
be smooth functions of quark mass and be the same for the quenched and
dynamical runs. 
The resulting minimum distances are strongly dependent on the quark masses,
with smaller minimum distances for smaller masses.
Most of this is due to the larger statistical errors at smaller quark mass,
which result in the excited state contributions disappearing into the noise
at shorter distance.  However, it is in part physical, since splittings between
the ground and excited states are larger for smaller quark masses.
Table \ref{MIN_D_TABLE} shows the minimum distances that we chose, and
the number of particles with each parity.

\section{Results}
\subsection{Pseudoscalar mesons}
\begin{figure}[tbp]
\epsfxsize=6.0in
\epsfysize=6.0in
\rule{0.1in}{0.0in}\hspace{1.0in}\epsfbox[0 0 4096 4096]{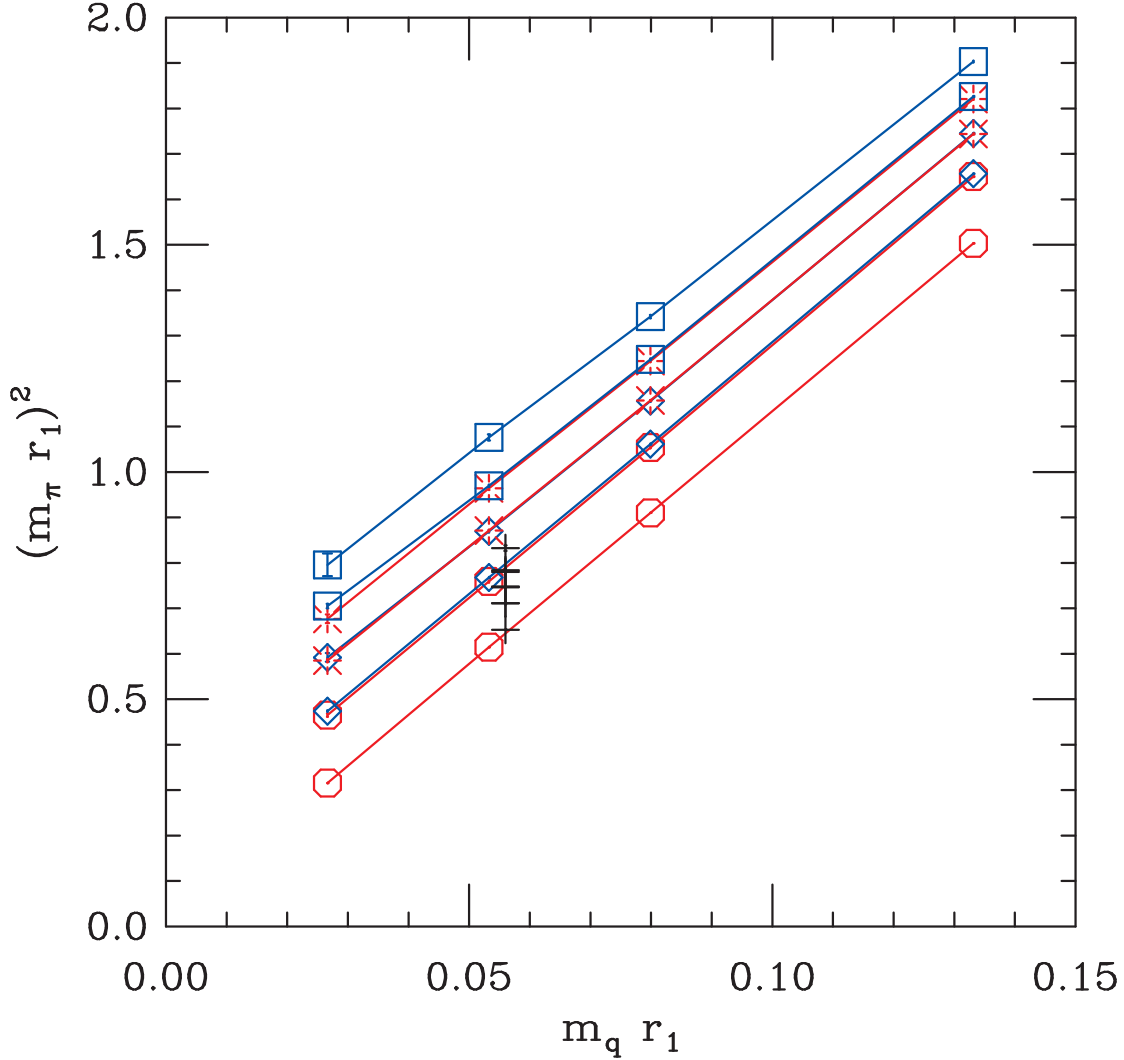}
\caption{
\label{MPISQ_VS_M_FIG}
Squared pseudoscalar meson masses versus quark masses.  These results
are from the quenched runs.  The octagons are the local pions
($\gamma_5\otimes\gamma_5$ and $\gamma_0\gamma_5 \otimes \gamma_0\gamma_5$),
the diamonds the distance one pions, the bursts the distance two pions, and
the squares the distance three pions.
The degeneracies predicted in Ref.~\protect\cite{PARTIAL_FLAVOR_SYM}
are clearly visible.  The lines are not fits; they simply connect the points.
The column of pluses is from the quenched $a \approx 0.09$ fm run, showing the
expected improvement in flavor symmetry with decreasing lattice spacing.
Note that the Goldstone pion changes very little when the lattice spacing is
decreased --- the non-Goldstone pions come down to join it.
}
\end{figure}

We calculated propagators for all eight flavor combinations
of the staggered quark pseudoscalar mesons.  These masses obey
the ``partial flavor symmetry restoration'' predicted by Lee and Sharpe\cite{PARTIAL_FLAVOR_SYM}
to very good accuracy.  Specifically, Ref.~\cite{PARTIAL_FLAVOR_SYM} predicts that the leading
order flavor symmetry breaking effects, which are order $a^2$, leave degeneracies
between pairs of pseudoscalar mesons for which $\gamma_0$ is replaced by $\gamma_i$
in the flavor structure.  For example, the local non-Goldstone pion, $\gamma_0\gamma_5 \otimes \gamma_0\gamma_5$
in the ``spin $\otimes$ flavor'' notation,
is degenerate with the distance one pion, $\gamma_5 \otimes \gamma_i\gamma_5$,
to this order.
Moreover, all of the squared pion masses should depend linearly on the
quark mass with the same slope to lowest order.
Figure~\ref{MPISQ_VS_M_FIG} shows this behavior for the quenched pion
masses for $am_{u,d} \le 0.05$.  The results for the full QCD runs are similar.
Since the local $\gamma_5\otimes\gamma_5$ pseudoscalar has the correct chiral
behavior (and the best scaling behavior), we will use this pseudoscalar in the rest of the analysis
unless we specifically indicate otherwise.
Results for the full QCD runs are similar, but the flavor symmetry
breaking is somewhat larger.

\begin{figure}[tbp]
\epsfxsize=6.0in
\epsfysize=6.0in
\rule{0.1in}{0.0in}\hspace{1.0in}\epsfbox[0 0 4096 4096]{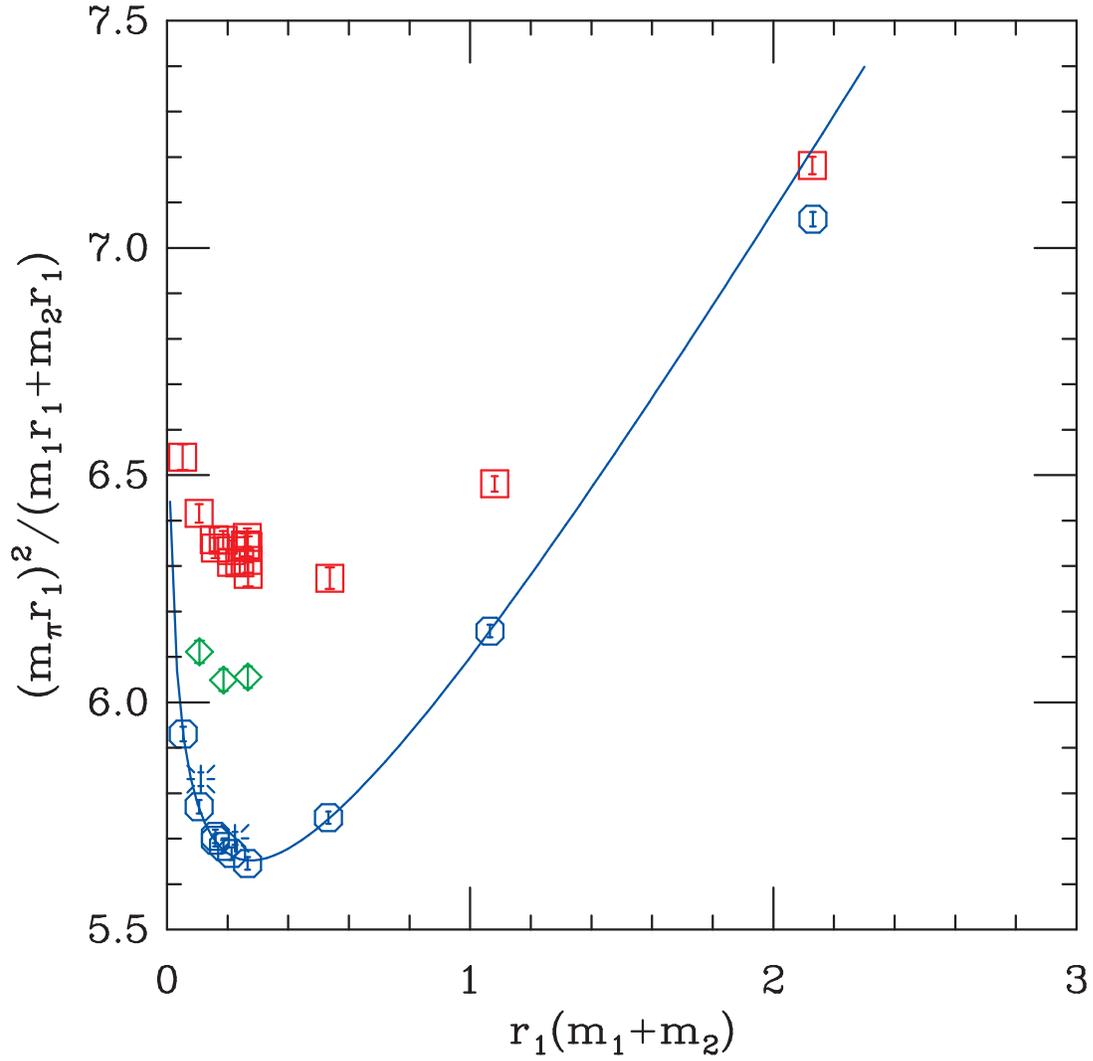}
\caption{
The squared pseudoscalar mass divided by the quark mass in units of $r_1$.
The octagons are the quenched $a\approx 0.13$ fm run, and the squares the
three flavor $a\approx 0.13$ fm runs.  The diamonds are the single two-flavor
run.
The bursts are from the quenched $a\approx 0.09$ fm run.
The fit is to the octagons (pions only) with
$r_1(m_1+m_2) < 1.4$, using the form in eq.~(\protect\ref{eq:QChPT}).
\label{MPISQ_OVER_R1_FIG}
}
\end{figure}

In Fig.~\ref{MPISQ_VS_M_FIG} the relation between the squared pseudoscalar
mass and the quark mass is clearly nearly linear.  The deviations from linearity
and the effect of the dynamical quarks can be exposed by plotting the squared
pseudoscalar masses divided by the quark mass, in Fig.~\ref{MPISQ_OVER_R1_FIG}.
This is essentially $\langle\psi\bar\psi\rangle  r_1/ f_\pi^2$ with an (unknown) renormalization factor.
This plot contains pseudoscalar mesons with both light and strange valence
quarks (pions, kaons and ``unmixed $s \bar s$'s'').
There is clearly a systematic difference between quenched and full QCD.
This difference increases with decreasing quark mass, and the two-flavor
point falls in between the quenched and three-flavor points.
The bursts among the quenched points are from the $10/g^2=8.4$, $a \approx 0.09$ fm
run, showing gratifying agreement with the $a \approx 0.13$ fm points.
Unfortunately, a coarser three flavor lattice, $a\approx 0.2$ fm,
shows a large effect, so we would not want to use much
coarser lattices in studying this effect.   
We do note that we expect scaling violations to be similar for the quenched and
dynamical theories, so it is an advantage to have runs with matched lattice
spacings.
The deviations from linearity of $m_\pi^2$ are similar in quenched and
full QCD.  The upturn for larger quark masses signals the beginning of the
transition to the heavy quark regime, where $m_\pi^2 \approx m_q^2$.  We do not
fully understand the shape of this plot for small quark mass.
There are several ways to interpret the difference between quenched and
dynamical results.  One could say that $\langle\bar\psi\psi\rangle/f_\pi^2$
is too small in the quenched approximation, or one could say that the quark mass
at which a desired $m_\pi/m_\rho$ is reached is larger in the quenched approximation
than in full QCD.
This second interpretation is consistent with CP-PACS results on the quark
masses, in which they find that the quark mass needed to reach a given value of
$m_\pi/m_\rho$ is smaller in two flavor QCD than in quenched QCD\cite{CPPACS_MASSEFFECT}.
(Indeed, one could even use this quantity as a length scale,
and conclude that $r_1$ is different in quenched and full QCD.)

\begin{figure}[tbp]
\epsfxsize=7.0in
\epsfysize=7.0in
\rule{0.1in}{0.0in}\hspace{-1.25in}\epsfbox[0 0 4096 4096]{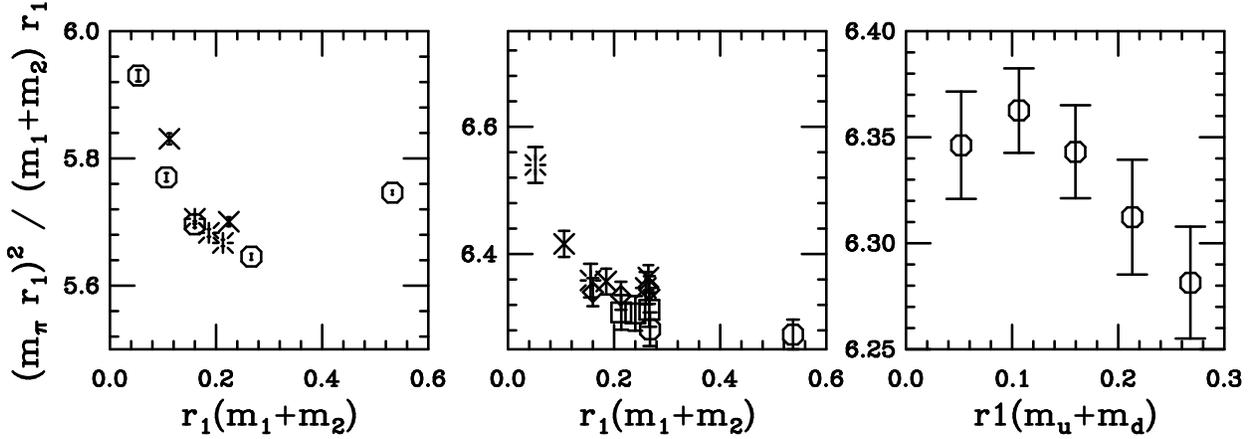}
\vspace{-4.0in}
\caption{
Details of $(m_{PS} r_1)^2/(r1(m_1+m_2))$.
The three panels are the quenched results, the dynamical results,
and the three flavor points with two strange valence quarks.
\label{MPISQ_DETAIL_FIG}
}
\end{figure}

The largest part of the error bars in Fig.~\ref{MPISQ_OVER_R1_FIG}
come from the uncertainty in $r_1$.   However, this uncertainty is common
to all of the points coming from the same set of lattices.  In particular,
all the $a \approx 0.13$ fm quenched points are correlated in this
respect, as are the three two-flavor points.  If we are interested in
the dependence of the pseudoscalar mass on the quark mass on a fixed lattice,
we may want to consider only the error from the determination of the meson
mass in units of $a$.  The left hand panel in Fig.~\ref{MPISQ_DETAIL_FIG}
shows $m_\pi^2 r_1^2 /(r_1(m_1+m_2))$ for the quenched calculation, including only
the error from $a m_{PS}$ and showing only the reasonably light mass points.
In this panel the
octagons are ``pions'', with $m_1=m_2=am_{u,d}$, and the bursts are ``kaons'',
with $am_2$ fixed at 0.05, which is approximately the physical value
of the strange quark mass. We see that this quantity is
dependent only on the sum of the quark masses to very good accuracy.
The center panel contains the same plot for the three-flavor runs, where
now $r_1$ is determined independently in each run.  In this
panel the octagons are runs with $am_{u,d}=am_s \ge 0.05$, the squares the run with
$am_{u,d}=0.04$, the diamonds from $am_{u,d}=0.03$, the crosses from 
$am_{u,d}=0.02$
and the bursts from $am_{u,d}=0.01$.  The three symbols for each of the runs
with $am_{u,d} < 0.05$ correspond to the ``pion'', with both valence quarks
light, the ``kaon'', with one light and one strange valence quark, and
an ``unmixed $s\bar s$'', with two valence quarks of mass approximately
equal to that of the strange quark, 
but no $q \bar q$ annihilation.  This graph is far from smooth, but most
of the scatter comes from the fact that each set of dynamical quark masses
has an independent uncertainty in $r_1$.
Note that each dynamical run (for example, the three bursts) shows
qualitatively the same behavior as the quenched case, with the light-light
pseudoscalar tending to a larger value.
An interesting question is how the sea quark mass affects the pseudoscalar
mass.  This can be investigated by looking at the points at
$a(m_1+m_2) = 0.1$,
which are the ``unmixed $s\bar s$'' points
with both valence quark masses equal to $0.05/a$.  In the right hand panel
of Fig.~\ref{MPISQ_DETAIL_FIG} we plot these points as a function of the light
quark mass.  There is a noticeable effect, with smaller light quark mass
producing larger $s\bar s$ mass.  The direction of this effect is consistent
with the smaller pseudoscalar masses in the two flavor and quenched calculations
seen in Fig.~\ref{MPISQ_OVER_R1_FIG}.

The selected pseudoscalar meson mass fits in units of the lattice spacing are
tabulated in table~\ref{PS_MASS_TABLE}.
In addition to the ``pions'', the table also contains fits
with one quark at about the strange quark mass and one lighter
quark, or ``kaons''.  For the two and three flavor runs we also tabulate
``unmixed $s \bar s$ mesons'', with two valence quarks
with $am_v=0.05$.

We have attempted to fit the results in Fig.~\ref{MPISQ_OVER_R1_FIG} 
to the forms predicted by chiral perturbation theory.  In the quenched
case, the behavior of $m_\pi^2$ as a function of quark mass
is derived in Refs.~\cite{CBandMG,Sharpe}. We use eqn.~(9) in \cite{CBandMG},
with the parameter $\alpha$, which is believed to be small, set equal to zero,
and the analytic correction term added to the chiral log.  For pions, 
then one has 
\begin{equation}
\label{eq:QChPT}
\frac{m_\pi^2}{m} = C(1 -\delta \log\left(\frac{Cm}{\Lambda^2}\right) + Km)\ ,
\end{equation}
where $m_1=m_2\equiv m$, $C$ and $K$ are constants, and the chiral scale
$\Lambda$ may be taken as the $\eta$ mass.  As shown in
Fig.~\ref{MPISQ_OVER_R1_FIG}, the fit to the $a=.13$ fm quenched data 
is good. It gives $\delta=0.061(3)$ (statistical error only), which is on the
low side but compatible with the range reported by CP-PACS \cite{CPPACS} 
and is in excellent agreement with the result of 
Bardeen {\it et al.} \cite{FermilabQChPT}, $\delta=0.065(13)$.

Unfortunately, our attempts to fit the 3-flavor pion data in Fig.~\ref{MPISQ_OVER_R1_FIG} to the
corresponding full QCD chiral form \cite{Leutwyler} have been unsuccessful
to date.  We can fit the 5 lowest-mass pions with reasonable
confidence level, but the coefficients of the analytic terms are
unreasonably large, and the fit misses the next lightest pion by a wide
margin.  If we try to fit the 6 lowest-mass pions, the fit has terrible
confidence level.  Finally, good fits can be obtained by introducing,
as an additional free parameter, an 
overall coefficient in front of the chiral logs.  
However, the value of that coefficient in the fit 
is much smaller than its predicted \cite{Leutwyler} value.  
We are continuing to study this puzzling situation.
Our current running at smaller lattice spacing may provide
additional insight here.

\newpage
\widetext
\narrowtext

\widetext
\narrowtext

\widetext
\narrowtext

\subsection{Vector mesons}

We calculated propagators for the two local vector mesons,
$\gamma_i\otimes\gamma_i$ (VT) and $\gamma_0\gamma_i\otimes\gamma_0\gamma_i$ (PV),
and two distance one vector mesons,
$\gamma_i\otimes{\bf 1}$ and $\gamma_0\gamma_i\otimes\gamma_0$.
Any flavor symmetry breaking among these mesons is smaller than
the statistical errors, so we simply quote results for the
local $\gamma_i\otimes\gamma_i$, or ``VT'' mesons.  Table~\ref{VEC_MASS_TABLE}
contains these masses in units of the lattice spacing.

\begin{figure}[tbp]
\epsfxsize=4.0in
\epsfysize=4.0in
\rule{0.1in}{0.0in}\hspace{1.0in}\epsfbox[0 0 4096 4096]{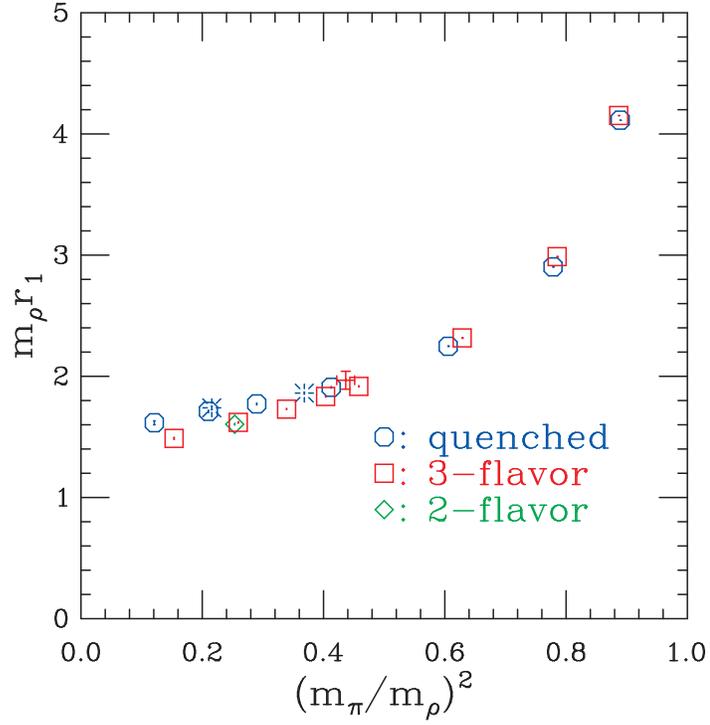}
\caption{
\label{MRHO_R1_FIG}
Vector meson masses in units of $r_1$.
The octagons are quenched results, the diamond a two-flavor
result, and the squares three-flavor.   The bursts
are quenched $a\approx 0.09$ fm points, and the fancy plus
an $a\approx 0.2$ fm three flavor point.
}
\end{figure}

\begin{figure}[tbp]
\epsfxsize=4.0in
\epsfysize=4.0in
\rule{0.1in}{0.0in}\hspace{1.0in}\epsfbox[0 0 4096 4096]{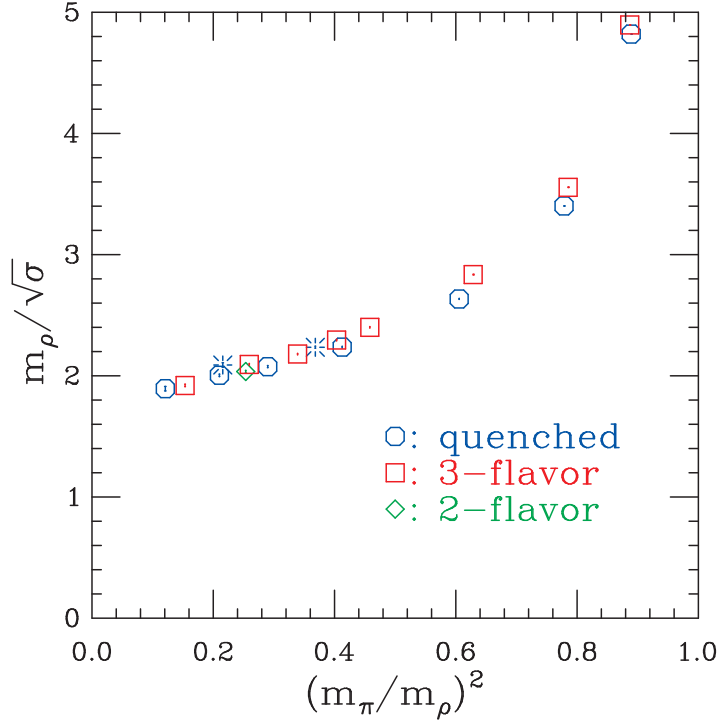}
\caption{
\label{MRHO_SIGMA_FIG}
The same as Fig.~\protect\ref{MRHO_R1_FIG}, except in
units of the string tension.
}
\end{figure}

In Figure~\ref{MRHO_R1_FIG} we plot the vector meson mass
in units of $r_1$ versus the squared pion/rho mass ratio.  In this plot
there is a clear difference between the quenched and dynamical
masses, with the full QCD vector mesons lying lower.  However,
the size of this effect depends on the length standard chosen,
as illustrated in Fig.~\ref{MRHO_SIGMA_FIG}, where the same
quantity is plotted in units of the string tension.  Of course,
the difference between these two plots simply arises from the
difference in $r_1 \sqrt{\sigma}$ plotted in Fig.~\ref{R1_SIGMA_FIG}.

\widetext
\narrowtext

\widetext
\narrowtext

In our calculations we used a mass of $am_s=0.05$ for the strange quark mass.
With the meson spectrum in hand, we can now go back and ask whether this
choice was exactly correct.  There are several quantities that we could choose to
define the strange quark mass, and, especially in the quenched calculation, they
will in general give different results.  Perhaps the most accessible quantity
is the ratio of the $s \bar s$ pseudoscalar to vector meson mass.  (Our ``$s \bar s$
pseudoscalar'' does not include $q \bar q$ annihilation diagrams, so it is
not the $\eta$ or $\eta^\prime$.)  We therefore tune this quantity using
$ m_{s \bar s} = \sqrt{ 2 m_K^2 - m_\pi^2 } = 686$ MeV and $m_\phi=1020$ MeV,
or $m_{PS}/m_V = 0.673$.

In the quenched spectrum with $am_q=0.05$ we have $m_{PS}/m_V = 0.643$, indicating
that $am_q=0.05$ is smaller than the desired strange quark mass.  To compute the
mass at which $m_{PS}/m_V = 0.673$, we can do a linear fit to the vector
meson mass as a function of quark mass, and combine that with a squared pseudoscalar
mass proportional to the quark mass to conclude that the quenched strange quark
mass defined by  $m_{PS}/m_V$ at this lattice spacing is $am_s=0.058$.
In contrast, for the dynamical runs $am_s=0.05$ is a fairly good estimate
of the strange quark mass, with $m_{PS}/m_V$ taking values between 0.670(2)
and 0.687(1).

\begin{figure}[tbp]
\epsfxsize=4.0in
\epsfysize=4.0in
\rule{0.1in}{0.0in}\hspace{1.0in}\epsfbox[0 0 4096 4096]{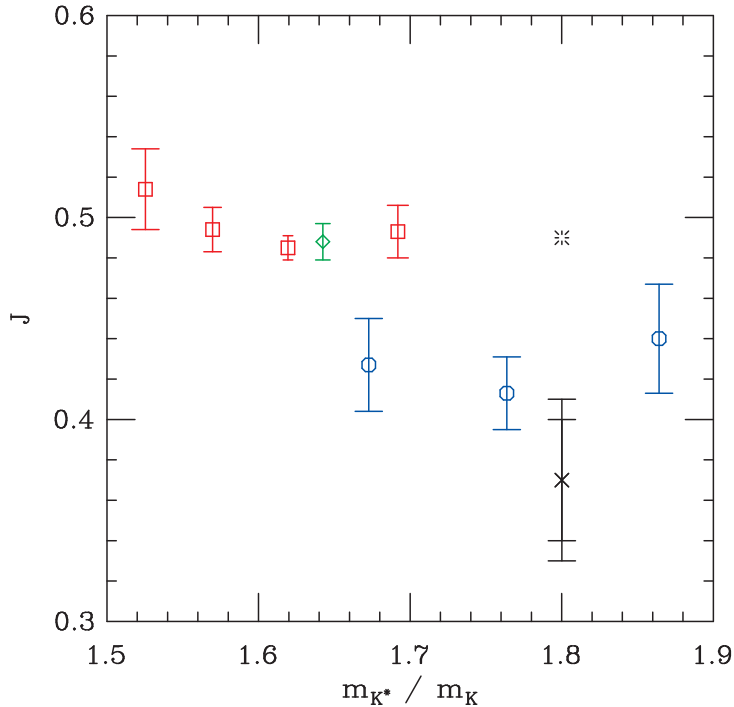}
\caption{
\label{J_FIG}
The mass ratio ``$J$'' in the quenched and full QCD calculations.
Squares are the three flavor results, and octagons are the quenched results.
The diamond is the two flavor run, using a non-dynamical strange quark
with mass $am_q=0.05$.
The burst is the real world value, and the cross is the UKQCD quenched
value. The smaller error bar on the cross is the statistical error, and
the larger the quoted systematic error.
}
\end{figure}

Differences between the quenched meson spectrum and the real world have
been observed by the UKQCD collaboration\cite{UKQCD_J}, and improvements
of the spectrum when dynamical quarks are included have been reported by
the CPPACS\cite{CPPACS_EFFECTS} and JLQCD\cite{JLQCD_EFFECTS} collaborations.
In particular, the UKQCD collaboration studied the quantity
\BE J = m_{K^*} \frac{\partial m_V}{\partial m_{PS}^2} \ \ \ ,\EE
where $m_V$ and $m_{PS}$ are the vector and pseudoscalar meson masses.
This quantity has the advantage of being relatively insensitive to the
quark masses, so that accurate tuning of the strange quark mass or
extrapolation of the masses to the chiral limit is not essential.
Of course,
to compare to experiment the derivative in this expression must be replaced
by a ratio of mass differences, and we choose
\BE \label{J_EQ}  J = m_{K^*} \frac{ \LP m_\phi-m_\rho \RP}{2 \LP m_K^2-m_\pi^2 \RP} \ \ \ .\EE
Here $m_\rho$ is the mass of the vector meson including two light quarks, etc.
We choose the $m_\phi-m_\rho$ mass difference because the statistical error in
$J$ is dominated by the error in the vector meson masses, and the larger difference
in $m_\phi-m_\rho$ relative to, say, $m_{K^*}-m_\rho$ leads to smaller statistical
errors on $J$.  Because all of the masses in Eq.~\ref{J_EQ} are correlated, we
use a jackknife analysis to compute the error on $J$.
Figure~\ref{J_FIG} shows the results for $J$ in quenched and three flavor QCD.
Following UKQCD, we plot this versus $m_{K^*}/m_K$, for which the real world
value is 1.8.  The burst is the real world value of
this definition of $J$ (0.49), and the cross is the value of $J$ found in the
UKQCD quenched simulations.  We see a clear effect of the sea quarks on this
quantity.
Indeed, any reasonable extrapolation of our data in  $m_{K^*}/m_K$ would pass
near the real world point.
Figure~\ref{J_FIG} also contains one point with two dynamical flavors.  This
point falls near the three flavor points, indicating that the dynamical
strange quark is less important than the two light quarks.
Although our quenched results are somewhat higher than the UKQCD
value, they are significantly below the experimental value.
The fact that the quenched points in this plot are to the right of the full
QCD points is largely due to the fact that the mass of $am_q=0.05$ used for
the quenched strange quark was too small, as discussed above.  If we use the
observed slopes of the quenched vector meson masses and squared pseudoscalar
masses as functions of quark mass to adjust these points to a quenched strange
quark mass of $am_s=0.058$, the main effect is to shift the quenched points
to the left.  In particular, the rightmost quenched point moves to
$m_{K^*}/m_K=1.78$, but moves up by only 0.004.

\widetext
\narrowtext

\subsection{P-wave mesons}

\begin{figure}[tbp]
\epsfxsize=6.0in
\epsfysize=6.0in
\rule{0.1in}{0.0in}\hspace{1.0in}\epsfbox[0 0 4096 4096]{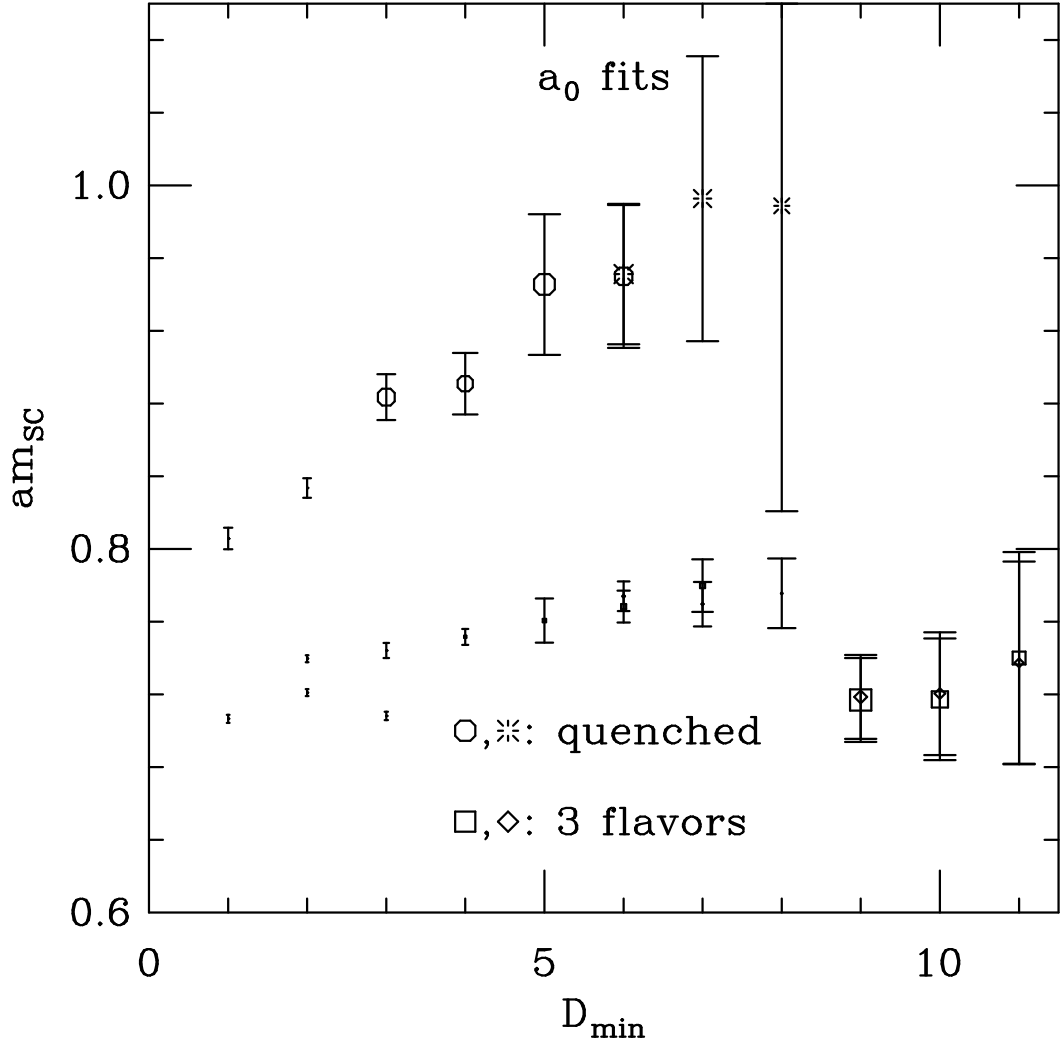}
\caption{
\label{DMIN_A0_FIG}
Fitted masses for the scalar ($a_0$) meson with quark mass $am_{u,d}=0.03$
as a function of the minimum distance included in the fit.  In this plot the
symbol size indicates the confidence level of the fit, with the symbol size
used in the legend corresponding to 50\%.
Here the octagons are quenched results using two $0^{-+}$ (pion) states and
one $0^{++}$ ($a_0$) state, while the bursts are quenched results with
one state of each parity.  The squares and diamonds are 2+1 flavor results
using two and one $0^{-+}$ states respectively.
}
\end{figure}

\begin{figure}[tbp]
\epsfxsize=6.0in
\epsfysize=6.0in
\rule{0.1in}{0.0in}\hspace{1.0in}\epsfbox[0 0 4096 4096]{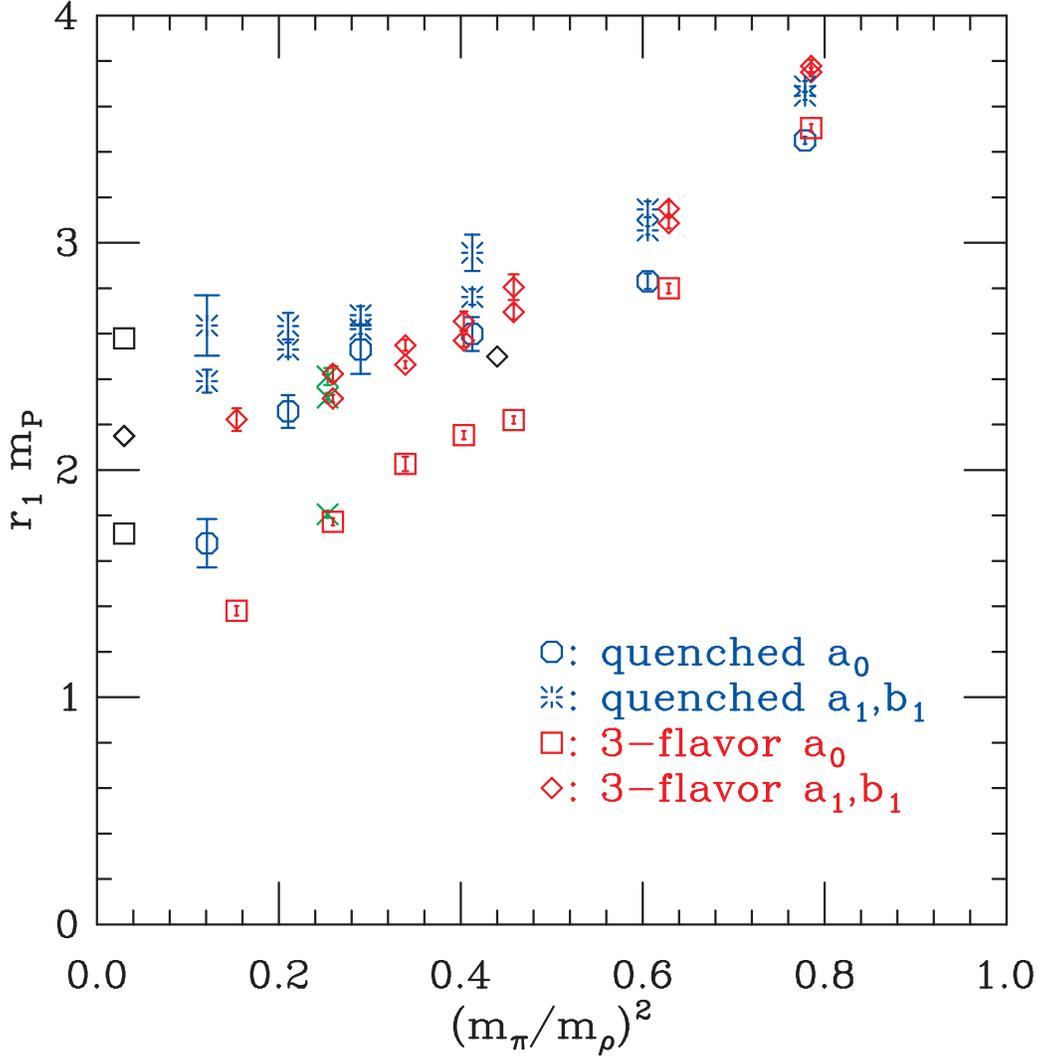}
\caption{
\label{PWAVES_FIG}
P-wave meson masses in units of $r_1$.
The bursts are the quenched pseudovector mesons ($a_1$ and $b_1$), and
the diamonds the full QCD pseudovectors.  Where the difference is significant
the $1^{+-}$ ($b_1$) is heavier than the $1^{++}$.  The octagons are the
quenched scalar ($a_0$), and the squares the full QCD scalar.
Crosses are the two flavor results.
The diamond at the physical value of $(m_\pi/m_\rho)^2$ is the
experimental $a_1$ and $b_1$ mass, and the two squares are experimental
$0^{++}$ ($a_0$) masses.
}
\end{figure}

Mesons with quantum numbers $J^{PC} = 0^{++}$, $1^{++}$ and
$1^{+-}$, which are P-wave mesons in the nonrelativistic limit, are found
as the oscillating ``parity partners'' of the $\pi$ and $\rho$ propagators
with Kogut-Susskind quarks.
Fitting these particles is difficult because of their larger mass.  In order
to get good fits, we need to allow two particles in the non-oscillating
components, since we typically find that the excited ``$\pi$'' or ``$\rho$''
state is comparable in mass to the lowest oscillating (P-wave) state.
Thus we find large errors, and must use small minimum distances to get
good fits. In addition, plateaus in the effective mass are short.
Figure~\ref{DMIN_A0_FIG} illustrates an example of a difficult but crucial
case.  This figure shows fits to the scalar ($a_0$) mass as a function
of the minimum distance included in the fit.
With these caveats, selected fits for the P-wave mesons are shown in 
Fig.~\ref{PWAVES_FIG} and Tables~\ref{A0_MASS_TABLE}, \ref{A1_MASS_TABLE} and \ref{B1_MASS_TABLE}.
This figure and these tables contain several interesting features.
For the $a_1$ and $b_1$ the full QCD runs give consistently better fits
than the quenched run, and smaller masses for the lighter quarks.
The $b_1$ is consistently slightly heavier than the $a_1$, although with
the difficulties in extracting these particles we would not want to 
make too much of this.  The diamond at the left in Fig.~\ref{PWAVES_FIG}
is the experimental value for the $a_1$ and $b_1$ masses.

\begin{figure}[tbp]
\epsfxsize=6.0in
\epsfysize=6.0in
\rule{0.1in}{0.0in}\hspace{1.0in}\epsfbox[0 0 4096 4096]{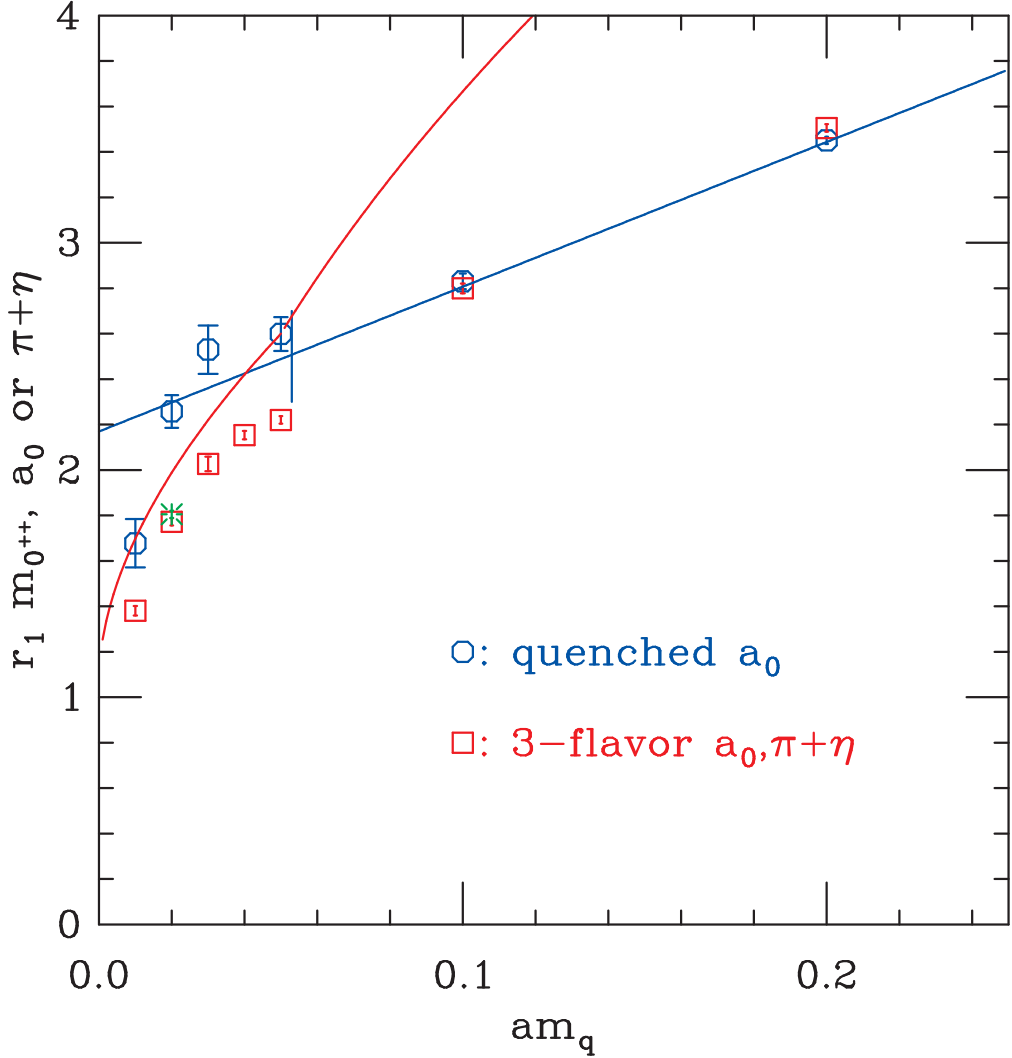}
\caption{
\label{A0_DECAY_FIG}
$0^{++}$ masses versus quark mass.
The lighest fitted energy in the scalar channel.  Octagons are
quenched results, squares are three flavor results, and the burst
is the two flavor run.
The straight line is a crude extrapolation of the heavy quark points.
The curved line is the $\pi + \eta$ mass estimate, as discussed in the
text.
The short vertical line marks the approximate quark mass where
the $a_0$ mass is twice the quenched pion mass.
}
\end{figure}

The scalar channel, ``$a_0$'', is clearly very different in the quenched and
full QCD runs.  For large quark masses there is no visible difference, but as
the quark mass is decreased the full QCD $0^{++}$ mass drops below
all the other masses.  For all but the lowest quark mass, the quenched $0^{++}$
is close to the other P-wave meson masses.  We ascribe the behavior of the
full QCD mass to the decay of the $a_0$ into $\pi+\eta$.  (Bose symmetry plus
isospin forbids decay into two pions.)  Figure~\ref{A0_DECAY_FIG} illustrates
this interpretation.  In the figure we plot the quenched and full $0^{++}$
masses versus quark mass.  The straight line in the graph is a fit to the
quenched mass for the heavier quarks, and represents the mass of a $q \bar q$
state.  The curved line with the kink at $am_q=0.05$ represents the mass
of $\pi+\eta$.  For $am_q \ge 0.05$ we used three degenerate quark flavors,
so the $\eta$ and $\pi$ are degenerate and this line is simply twice the pion
mass.  For $am_q<0.05$ we don't have direct information on the $\eta$ mass, so
we use the Gell-man--Okubo formula written in terms of an ``unmixed $s\bar s$''
mass (just our pseudoscalar mass at $am_q=0.05$).
\BE m_\eta^2 = \LP m_\pi^2 + 2 m_{s \bar s}^2 \RP/3 \EE
In the quenched case the $a_0$ mesons can couple to two-meson
states through a "hairpin diagram" on one of the meson lines.
Such diagrams, like Fig.~1(b)  in Ref.~\cite{CBandMG96}, can
behave like powers of $t$ times $e^{-2 m_\pi t}$ and therefore
masquerade as a light $a_0$ when $2 m_\pi < m_{a_0}$.  This may
explain the lightest quark mass quenched point.
In this analysis we used the $a_0$ from the local source, or the
``${\bf 1} \otimes {\bf 1}$'' operator, which gave the best signal.  Within the
very large statistical errors, we saw relatively large breaking of flavor
symmetry among the different $a_0$ channels, with some evidence that this
reflects the masses of the different lattice pseudoscalars to which the
various $a_0$'s should couple.

\widetext
\narrowtext

\widetext
\narrowtext

\widetext
\narrowtext

\subsection{Baryons}

We have evaluated propagators for baryons using the ``corner wall'' source
for both degenerate and nondegenerate quarks, using a pointlike sink operator
with all three quarks on the same lattice site.  With nondegenerate quarks,
the lightest states in this channel for zero, one or two strange quarks
are the $N$, the $\Lambda$ and the $\Xi$ respectively.
However, since we took no special measures to make the operator with one
strange quark and two light quarks orthogonal to the $\Sigma$, these propagators
undoubtedly contain contamination from a nearby $\Sigma$.   
In order to get the decuplet baryons, we followed Ref.~\cite{LANL_SPECTRUM},
using a wall source on every spatial site, and using the operator in Eq.~6.3
of Ref.~\cite{GOLTERMAN_BARYONS} for the $\Delta$.  This is necessary because
the corner wall source does not overlap this $\Delta$ operator.
As a byproduct of the calculation of the decuplet mass, we obtain nucleon
propagators from even site wall sources and a wall source containing all
sites.  These propagators are generally noisier than the corner wall source
propagators, and the plateau in the effective mass occurs at larger distances.
Therefore we generally use the corner source propagators.  However, if one
uses our fitting procedures on the even-wall or full-wall propagators,
one invariably selects a smaller mass than from the corner-wall source
propagators.  This situation is illustrated by Fig.~\ref{NUC_DMIN_FIG}, which
show mass fits to these two propagators as a function of the minimum distance
included in the fit.  In this figure the corner source propagators reach a plateau
earlier and with smaller error bars.  However, there are perfectly acceptable fits
to the full wall source propagators giving masses significantly smaller than
the corner source values.
Perhaps the only good thing we can say about this situation is that the effect is similar
in the quenched and dynamical runs, so as long as we are careful to make the same choices
in both cases, we can investigate the effects of sea quarks on the spectrum while
taking the statistical errors at face value.
The results of the fits that we selected are listed
in Tables~\ref{BARYON_MASS_TABLE} and \ref{DELTA_MASS_TABLE}.

The nucleon to rho mass ratio, or ``Edinburgh plot'', has long been used as a way
of displaying lattice spectrum results.  This ratio is known to be sensitive to
lattice spacing, lattice volume and quark masses.  As mentioned above, we do not
address the issues of continuum extrapolation and chiral extrapolation in this
paper.  However, since our quenched and dynamical lattices are matched in lattice
spacing and physical size, we are well positioned to ask if effects of dynamical
quarks show up in this ratio.
In Fig.~\ref{APE_BOTH_FIG} we show a variant of the Edinburgh plot, the ``APE plot'',
where $(m_\pi/m_\rho)^2$ is used as the abscissa, so that for small quark masses
the abscissa is proportional to the quark mass.
Most of the points on this plot are from the $a \approx 0.13$ fm matched lattice runs.
It can be seen that there are no significant differences between the quenched and
three flavor runs.  The single two flavor point lies slightly above the trend, although
this is probably not significant.
This agreement between $m_N/m_\rho$ for quenched and full QCD is in apparent conflict
with our extrapolations of the conventional action\cite{MILC_DIFFERENT}.
The discrepancy, which may be due to residual discretization
effects in one or both calculations, is under study.
We are hopeful that currently running three flavor simulations 
at finer lattice spacing will shed new light here.
Figure~\ref{APE_BOTH_FIG} also contains a point from a coarser lattice three flavor run and a preliminary
point from a finer lattice quenched run ($10/g^2=8.4$, $28^3\times 96$ lattice, $a \approx 0.09$
fm)  These two points suggest that when we are in a position to do a continuum extrapolation
the continuum results will be lower.

It is interesting to compare these results to the conventional Kogut-Susskind quark
action.  Figure~\ref{APE_QUENCH_FIG} shows the improved action results together
with conventional action results at $6/g^2 = 5.7$, $5.85$. $6.15$\cite{MILC_QUENCHED}
 and $6.5$\cite{KIM_AND_OHTA},
which correspond to lattice spacings of about $0.16$, $0.12$, $0.07$ and 0.043 fm
respectively.  While a continuum extrapolation will be deferred until the 0.1 fm
runs are completed, we can see in this plot that the improved action at 0.13 fm 
gives results similar to the conventional action at $a=0.07$ fm.

Just as for the mesons, one of the biggest problems in comparing baryon masses to
the real world is the need for an extrapolation in quark mass.  For the mesons the
quantity $J$ has the nice feature that it is only minimally sensitive to this extrapolation.
It is tempting to try to construct similar quantities for the baryons.  This suggests looking
at the $\Omega^-$, the $sss$ decuplet baryon.  More generally, we could plot the
mass of the decuplet baryon as a function of $(m_{PS}/m_V)^2$ to produce a variant
of the Edinburgh plot which has the nice feature that there are two experimental points, 
one for $m_{\Omega^-}/m_\phi$ and another for $m_\Delta/m_\rho$.  Unfortunately, because
the decuplet masses are hard to determine on the lattice, the errors are rather large.  The result
of this exercise is in Fig.~\ref{APE_STRANGE_FIG}.
Although the error bars and the scatter among the points are large, the overall
trend of this plot is encouraging.  It is premature to say whether this plot
shows real differences between full and quenched QCD.

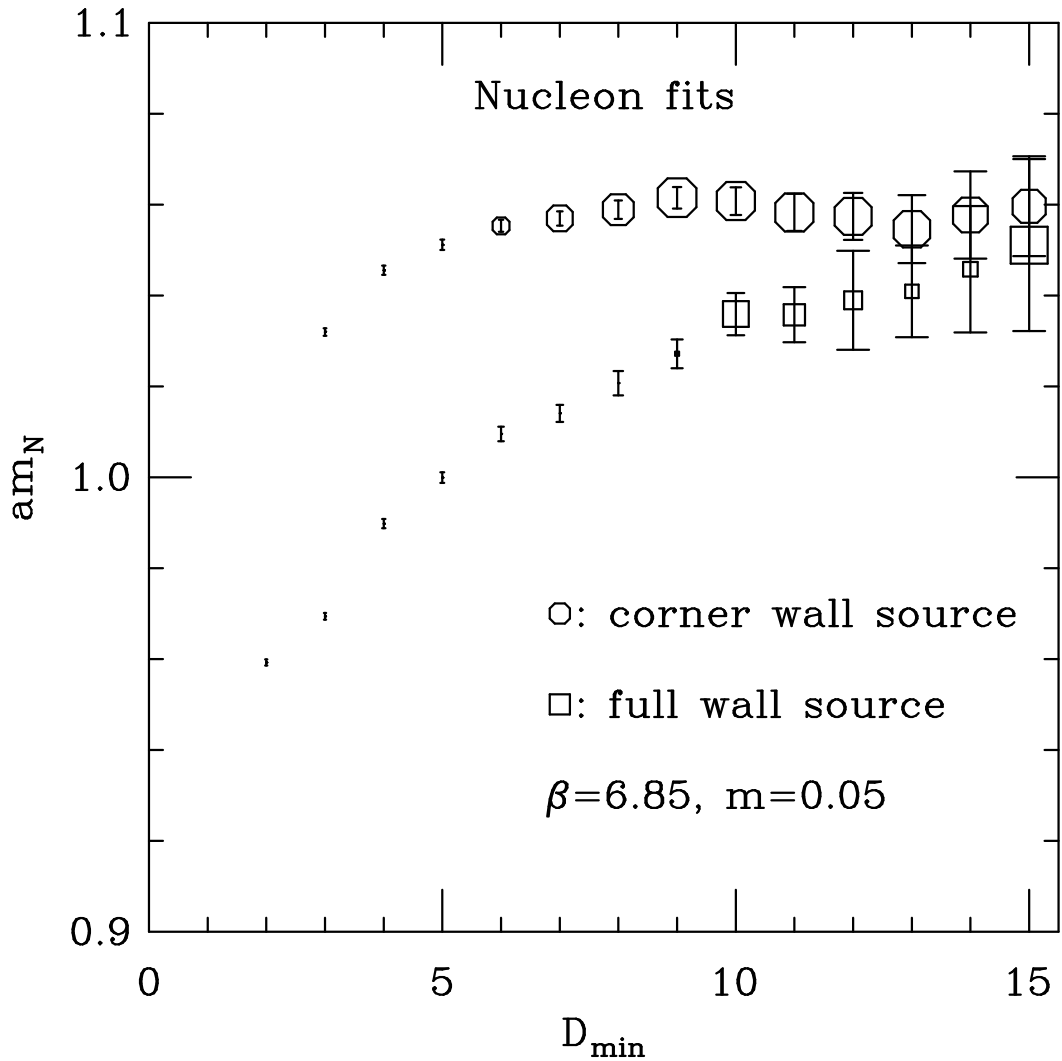
\widetext
\begin{figure}[tbp]
\epsfxsize=6.0in
\epsfysize=6.0in
\rule{0.1in}{0.0in}\hspace{1.0in}\epsfbox[0 0 4096 4096]{dmin_nuc_b685.ps}
\caption{
\label{NUC_DMIN_FIG}
Mass fits to nucleon propagators as a function of the minimum distance included
in the fit.  These fits are from the three flavor run with $10/g^2=6.85$ and $am_q=0.05$,
and are fairly typical.   Here the octagons are from propagators with a ``corner wall''
source and point sink on all sites, and the squares are from the ``full wall'' source
with a point sink on even sites only.  In this plot, the symbol size at each point
is proportional to the confidence level of the fit, on a scale where the symbols in
the legend correspond to 50\% confidence.   For this quark mass we used $D_{min}=12$
for our quoted mass.
}
\end{figure}
\narrowtext

\widetext
\begin{figure}[tbp]
\epsfxsize=6.0in
\epsfysize=6.0in
\rule{0.1in}{0.0in}\hspace{1.0in}\epsfbox[0 0 4096 4096]{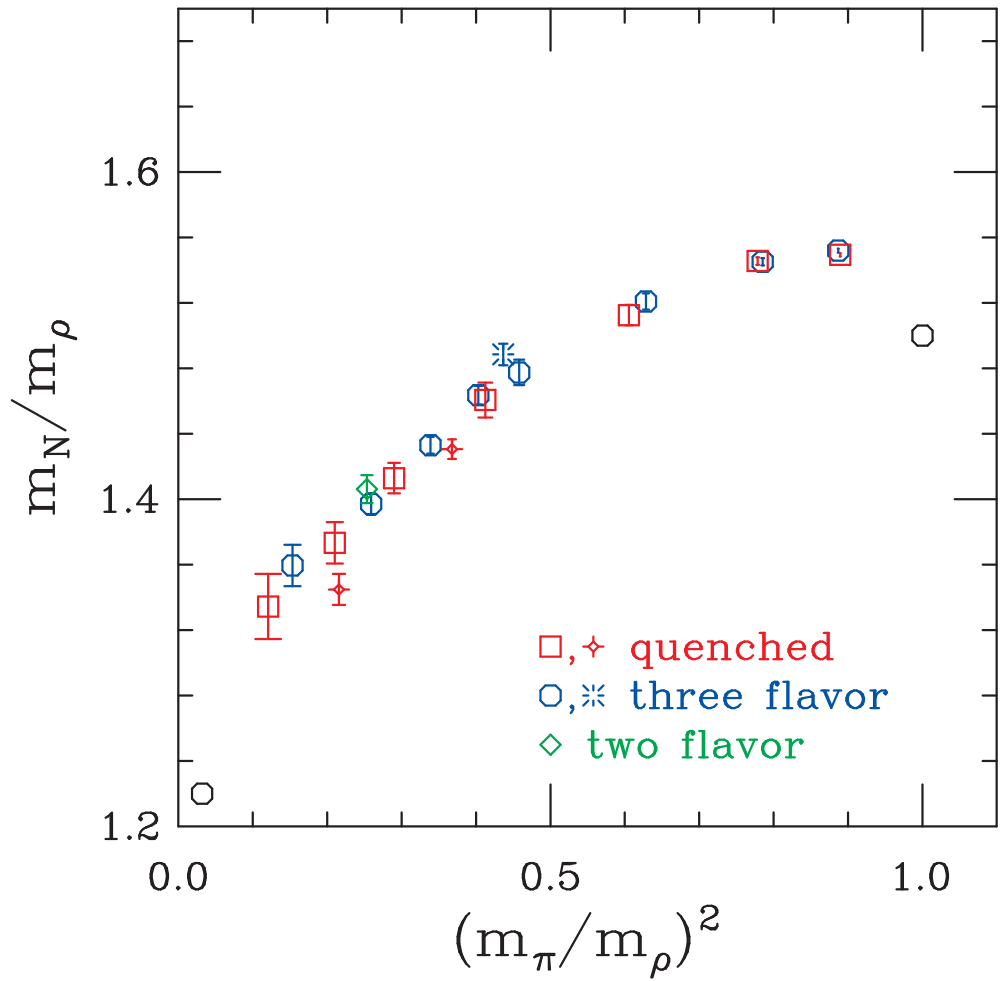}
\caption{
\label{APE_BOTH_FIG}
The nucleon to rho mass ratio in quenched and full QCD.  The squares, diamond and octagons
are the $a \approx 0.13$ fm matched lattice runs with zero, two and three flavors
respectively.  The fancy diamonds below the other points are preliminary quenched
points at $a \approx 0.09$ fm, and the burst lying above the trend is a coarse lattice
three flavor run at $a \approx 0.2$ fm.
The octagon at the left is the physical value, and the octagon at the right is the
trivial infinite quark mass value.
}
\end{figure}
\narrowtext

\widetext
\begin{figure}[tbp]
\epsfxsize=6.0in
\epsfysize=6.0in
\rule{0.1in}{0.0in}\hspace{1.0in}\epsfbox[0 0 4096 4096]{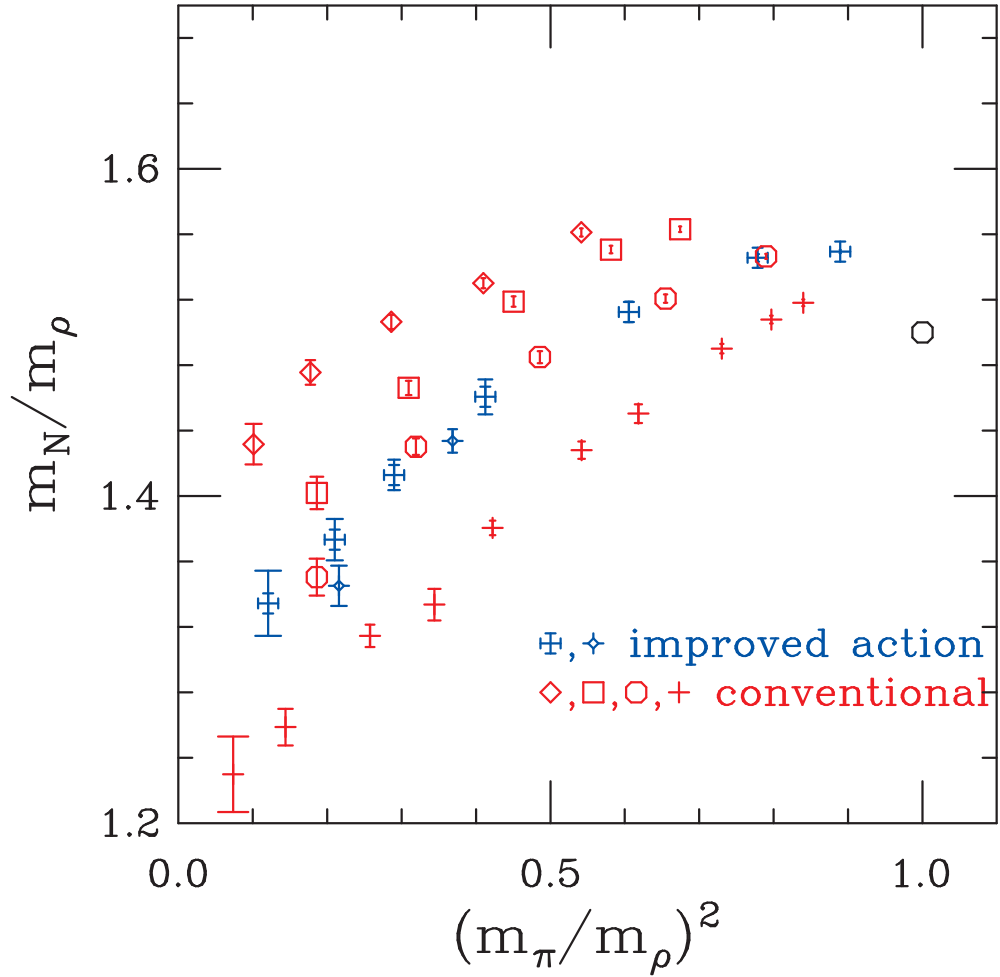}
\caption{
\label{APE_QUENCH_FIG}
The nucleon to rho mass ratio in quenched QCD with conventional and
improved Kogut-Susskind quark action.
The diamonds, squares, octagons and
pluses are from the conventional action with lattice spacings of
$0.16$, $0.12$, $0.07$ and $0.043$ fm respectively.
The decorated pluses are the improved action results with $a \approx 0.13$ fm, and
the fancy diamonds are preliminary improved action results at $a \approx 0.09$ fm.
}
\end{figure}
\narrowtext

\widetext
\begin{figure}[tbp]
\epsfxsize=6.0in
\epsfysize=6.0in
\rule{0.1in}{0.0in}\hspace{1.0in}\epsfbox[0 0 4096 4096]{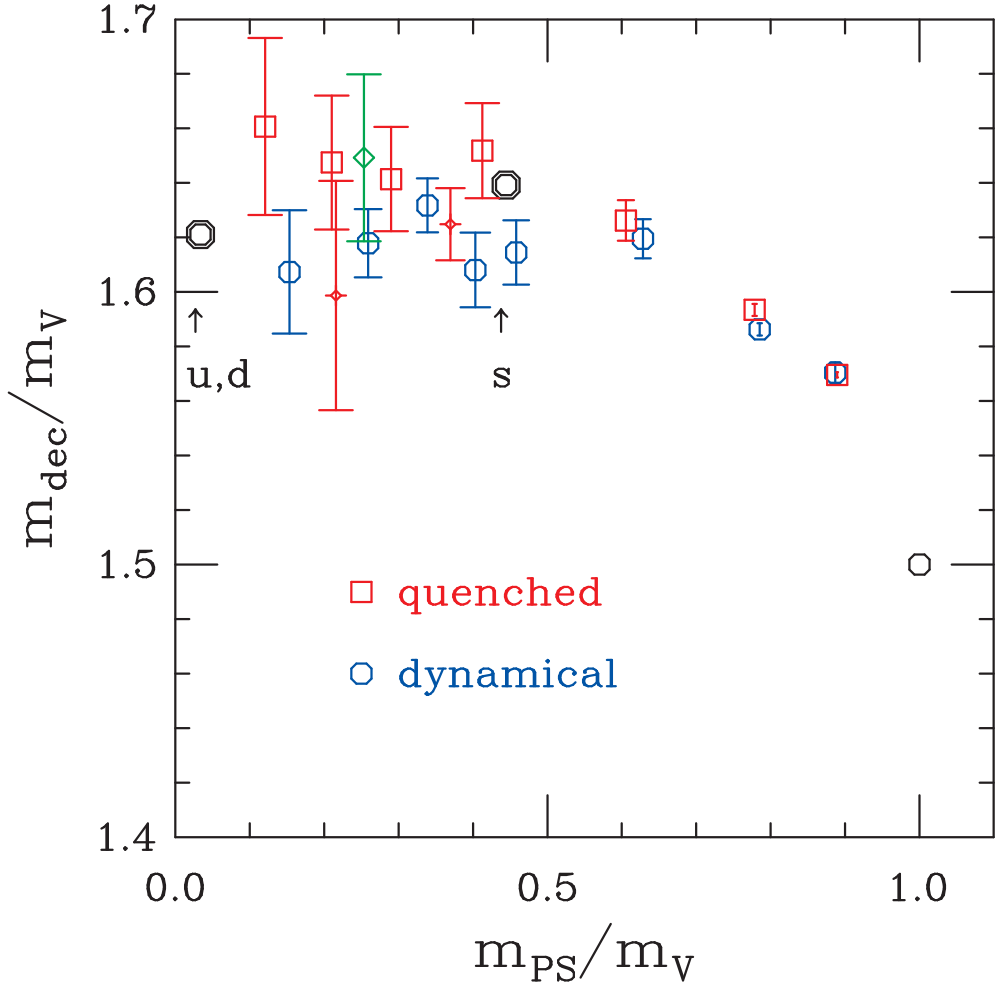}
\caption{
\label{APE_STRANGE_FIG}
The decuplet baryon to vector meson mass ratio at $a \approx 0.13$ fm.
Squares are the $a\approx 0.13$ fm quenched runs, and octagons the
$a\approx 0.13$ fm three flavor runs.  The single diamond is the
$a\approx 0.13$ fm two flavor run.
The bold octagons without error bars are $m_{\Omega^-}/m_\phi$ and
$m_\Delta/m_\rho$.
}
\end{figure}
\narrowtext

\subsection{Nonzero momentum}

As a check on the quality of the dynamics in our simulations, we calculated the
energies of a few nonzero momentum mesons, namely the Goldstone pion with momenta
$\frac{2\pi}{L}\LP 0,0,1 \RP$ and $\frac{2\pi}{L}\LP 0,1,1 \RP$ and the $\gamma_z \otimes
\gamma_z$ rho with momentum $\frac{2\pi}{L}\LP 0,0,1 \RP$.  
We compare these energies to the ideal dispersion relation by tabulating the ``speed
of light'' in Table~\ref{C_TABLE}.
\BE c^2 = \frac{ E(\vec k) - E(\vec 0) }{k^2} \EE
We see that the dispersion relation is generally very good, but with 
noticeable deviations from one for the heavier mesons.
The results are similar in the quenched and full QCD calculations,
as can be seen by comparing lines with the same valence quark
mass in Table~\ref{C_TABLE}.

\section{Conclusion}

We have used simulations with three flavors of dynamical quarks and a
quenched simulation on lattices with matched lattice spacing and physical
size to isolate effects of the sea quarks on the hadron spectrum and on
the static quark potential.  This was done with a Symanzik improved gauge
action and an improved Kogut-Susskind quark action to make the effects
of the nonzero lattice spacing as small as practical.  Effects of the
sea quarks are clearly visible in the static potential.  The mesonic
mass ratio $J$ is much closer to the experimental value when dynamical
quarks are included.  The $a_0$ meson couples strongly to two meson
states, as expected when sea quarks are included.

Several aspects need further study.  While the quenched pion mass
has the expected form, we have not understood the dependence of
the three flavor pion mass on the quark masses.  It would be
very nice to be able to extract excited state masses, especially
in the avoided level crossing of the $a_0$, and to extract
decay rates to be compared to experiment.   Although we have
minimized the lattice artifacts by using an improved action, an
empirical investigation of these effects is necessary.  We are beginning
a series of simulations with $a \approx 0.09$ fm to investigate these
effects.

\section*{acknowledgements}

Computations were done on the Origin 2000 clusters at LANL,
on the T3E and SP3 at NERSC, on the T3E, SP2 and SP3
at SDSC, on the NT and Linux clusters and Origin 2000 at NCSA, on the Origin
2000 at BU, on the Roadrunner and Los Lobos clusters at UNM
and on the PC clusters at NERSC, Indiana University and the University of Utah.
This work was supported by the U.S. Department of Energy under contracts
DOE -- DE-FG02-91ER-40628,	
DOE -- DE-FG03-95ER-40894,	
DOE -- DE-FG02-91ER-40661,	
DOE -- DE-FG02-97ER-41022 	
and
DOE -- DE-FG03-95ER-40906 	
and National Science Foundation grants
NSF -- PHY99-70701 		
and
NSF -- PHY97--22022.    	


\begin{table}
\begin{center}
\setlength{\tabcolsep}{1.5mm}
\begin{tabular}{|l|l|l|l|l|l|l|}
$am_{u,d}$ / $am_s$  & \hspace{-1.0mm}$10/g^2$  & $u_0$ & res. & $\epsilon$ & lats. & $a/r_1$ \\
\hline                             
quenched       & 8.00  & 0.8879 & na			& na    & 408   & 0.3762(8) \\
\hline                             
0.02  / na     & 7.20  & 0.8755 & $1\times 10^{-4}$	& 0.013 & 370 & 0.3745(14) \\
\hline                             
0.40  / 0.40   & 7.35  & 0.8822 & $2\times 10^{-5}$	& 0.03  & 332   & 0.3766(10) \\
0.20  / 0.20   & 7.15  & 0.8787 & $5\times 10^{-5}$	& 0.03  & 341   & 0.3707(10) \\
0.10  / 0.10   & 6.96  & 0.8739 & $5\times 10^{-5}$	& 0.03  & 339   & 0.3730(14) \\
0.05  / 0.05   & 6.85  & 0.8707 & $1\times 10^{-4}$	& 0.02  & 425   & 0.3742(15) \\
0.04  / 0.05   & 6.83  & 0.8702 & $5\times 10^{-5}$	& 0.02  & 351	& 0.3765(14) \\
0.03  / 0.05   & 6.81  & 0.8696 & $5\times 10^{-5}$	& 0.02  & 564	& 0.3775(12) \\
0.02  / 0.05   & 6.79  & 0.8688 & $1\times 10^{-4}$	& 0.0133 & 484	& 0.3775(12) \\
0.01  / 0.05   & 6.76  & 0.8677 & $1\times 10^{-4}$	& 0.00667& 407	& 0.3852(14) \\   
\end{tabular}
\caption{Parameters of the $a=0.13$ fm simulations}
\label{RUN_TABLE}
\end{center}
\end{table}

\begin{table}
\begin{center}
\setlength{\tabcolsep}{1.5mm}
\begin{tabular}{|c|c|cccccccc|}
  &  & 0.01 & 0.02 & 0.03 & 0.04 & 0.05 & 0.1 & 0.2 & 0.4 \\
H		& fit & & & & & & & & \\
$\pi$		& 1,0 & 18 & 18 & 18 & 18 & 18 & 18 & 18 & 18 \\
$K$		& 1,0 & 14 & 14 & 14 & 14 &    &    &    &     \\
$s \bar s$	& 1,0 & 14 & 14 & 14 & 14 &    &    &    &     \\
$a_0$		& 2,1 & 4 & 4 & 5 & 5 & 6 & 6 & 6 & 6 \\
$\rho$		& 1,1 & 6 & 7 & 8 & 9 & 9 & 10 & 12 & 14 \\
$K^*$		& 1,1 & 8 & 8 & 9 & 9 &   &    &    &     \\
$\phi$		& 1,1 & 8 & 8 & 9 & 9 &   &    &    &     \\
$a_1$		& 2,1 & 4 & 4 & 4 & 5 & 5 & 5 & 6 & 7 \\
$b_1$		& 2,1 & 4 & 4 & 4 & 5 & 5 & 5 & 6 & 7 \\
$N$		& 1,1 & 7 & 8 & 9 & 10 & 12 & 14 & 16 & 18 \\
$\Lambda$	& 1,1 & 7 & 7 & 8 & 9 &   &    &    &     \\
$\Xi$		& 1,1 & 7 & 7 & 8 & 9 &   &    &    &     \\
$\Delta$	& 1,1 & 3 & 4 & 5 & 6 & 7 & 8 & 9 & 10 \\
\end{tabular}
\caption{Minimum distances used in propagator fits.
With the exception of the $\Delta$, these hadrons are obtained from
the ``corner'' source.
The top row is the light quark mass.  The second column shows the
type of fit used, where the two numbers are the number of simple
exponentials included and the  number of oscillating contributions
included.  For example, a fit of type ``2,1'' would include two
particles with one parity and one particle with the opposite parity.
Hadrons with nondegenerate valence quarks, such as the $K$, were
computed only for $m_{u,d} < m_s$.
}
\label{MIN_D_TABLE}
\end{center}
\end{table}

\begin{table}
\begin{center}
\begin{tabular}{|llllll|}
$am_{valence}$ & $am_{sea}$ & $am_{PS}$ & range & $\chi^2/D$ & conf. \\
\hline

0.4 ($\pi$)	& $\infty$	& 1.45664(12)	& 18-31	& 13/12	& 0.39 \\
0.2 ($\pi$)	& $\infty$	& 0.96166(16)	& 18-31	& 14/12	& 0.32 \\
0.1 ($\pi$)	& $\infty$	& 0.65693(19)	& 18-31	& 14/12	& 0.28 \\
0.05 ($\pi$)	& $\infty$	& 0.46043(20)	& 18-31	& 15/12	& 0.25 \\
0.03 ($\pi$)	& $\infty$	& 0.35825(16)	& 18-31	& 21/12	& 0.05 \\
0.02 ($\pi$)	& $\infty$	& 0.29440(17)	& 18-31	& 23/12	& 0.026 \\
0.01 ($\pi$)	& $\infty$	& 0.21104(16)	& 18-31	& 26/12	& 0.01 \\
0.03/0.05 ($K$)	& $\infty$	& 0.41261(19)	& 14-31	& 26/16	& 0.06 \\
0.02/0.05 ($K$)	& $\infty$	& 0.38650(20)	& 14-31	& 25/16	& 0.07 \\
0.01/0.05 ($K$)	& $\infty$	& 0.35852(23)	& 14-31	& 23/16	& 0.13 \\
\hline

0.02 ($\pi$)    & 0.02  & 0.30258(22)   & 18-31 & 9.4/12        & 0.67 \\
0.02/0.05 ($K$) & 0.02  & 0.39823(25)   & 14-31 & 33/16 & 0.0076 \\
0.05 ($s\bar s$)        & 0.02  & 0.47623(25)   & 14-31 & 27/16 & 0.047 \\
\hline

0.4 ($\pi$)     & 0.4   & 1.46932(17)   & 18-31 & 5.3/12        & 0.95 \\
0.2 ($\pi$)     & 0.2   & 0.97930(25)   & 18-31 & 8.3/12        & 0.76 \\
0.1 ($\pi$)     & 0.1   & 0.68332(24)   & 18-31 & 16/12 & 0.17 \\
0.05 ($\pi$)    & 0.05  & 0.48422(21)   & 18-31 & 26/12 & 0.011 \\
0.04 ($\pi$)    & 0.04/0.05     & 0.43507(27)   & 18-31 & 14/12 & 0.31 \\
0.03 ($\pi$)    & 0.03/0.05     & 0.37787(18)   & 18-31 & 15/12 & 0.25 \\
0.02 ($\pi$)    & 0.02/0.05     & 0.31125(16)   & 18-31 & 16/12 & 0.21 \\
0.01 ($\pi$)    & 0.01/0.05     & 0.22446(22)   & 18-31 & 14/12 & 0.27 \\
0.04/0.05 ($K$) & 0.04/0.05     & 0.46141(27)   & 14-31 & 18/16 & 0.31 \\
0.03/0.05 ($K$) & 0.03/0.05     & 0.43613(19)   & 14-31 & 26/16 & 0.052 \\
0.02/0.05 ($K$) & 0.02/0.05     & 0.40984(21)   & 14-31 & 19/16 & 0.28 \\
0.01/0.05 ($K$) & 0.01/0.05     & 0.38334(29)   & 14-31 & 25/16 & 0.072 \\
0.05 ($s\bar s$)        & 0.04/0.05     & 0.48659(27)   & 14-31 & 18/16 & 0.35 \\
0.05 ($s\bar s$)        & 0.03/0.05     & 0.48796(18)   & 14-31 & 28/16 & 0.035 \\
0.05 ($s\bar s$)        & 0.02/0.05     & 0.49009(20)   & 14-31 & 23/16 & 0.12 \\
0.05 ($s\bar s$)        & 0.01/0.05     & 0.49443(25)   & 14-31 & 19/16 & 0.26 \\
\end{tabular}
\caption{Pseudoscalar meson masses.
Quenched results are first, followed by the single two-flavor
run, followed by the three flavor runs.
The first column is the valence quark mass(es), and the
second column the sea quark mass or masses.
The particle name in the first column is intended as a mnemonic.  Here
``$\pi$'' indicates valence quark mass equal to the lighter dynamical
quarks, or degenerate in the quenched case. ``$K$'' indicates one valence
quark equal to the light dynamical quarks and one at about $m_s$, while
``$s \bar s$''  indicates two valence quarks with mass about $m_s$, in
a flavor nonsinglet state.
The remaining columns are the hadron mass, the time range for the
chosen fit, $\chi^2$ and number of degrees of freedom for the
fit, and the confidence level of the fit.
}
\label{PS_MASS_TABLE}
\end{center}
\end{table}

\begin{table}
\begin{center}
\begin{tabular}{|llllll|}
$am_{valence}$ & $am_{sea}$ & $am_{V}$ & range & $\chi^2/D$ & conf. \\
\hline

0.4 ($\rho$)	& $\infty$	& 1.5446(2)	& 14-31	& 18/14	& 0.19 \\
0.2 ($\rho$)	& $\infty$	& 1.0900(5)	& 12-31	& 19/16	& 0.27 \\
0.1 ($\rho$)	& $\infty$	& 0.8443(11)	& 10-28	& 21/15	& 0.15 \\
0.05 ($\rho$)	& $\infty$	& 0.7168(21)	& 9-27	& 10/15	& 0.8 \\
0.03 ($\rho$)	& $\infty$	& 0.6653(26)	& 8-23	& 6.9/12	& 0.86 \\
0.02 ($\rho$)	& $\infty$	& 0.6422(30)	& 7-21	& 7/11	& 0.8 \\
0.01 ($\rho$)	& $\infty$	& 0.6070(60)	& 6-16	& 10/7	& 0.19 \\
0.03/0.05 ($K^*$)	& $\infty$	& 0.6910(30)	& 9-24	& 13/12	& 0.38 \\
0.02/0.05 ($K^*$)	& $\infty$	& 0.6820(30)	& 8-21	& 10/10	& 0.42 \\
0.01/0.05 ($K^*$)	& $\infty$	& 0.6680(40)	& 8-21	& 11/10	& 0.34 \\
\hline

0.02 ($\rho$)	& 0.02	& 0.6009(23)	& 7-23	& 17/13	& 0.2 \\
0.02/0.05 ($K^*$)	& 0.02	& 0.6532(23)	& 8-27	& 22/16	& 0.14 \\
0.05 ($\phi$)	& 0.02	& 0.7003(14)	& 8-29	& 23/18	& 0.21 \\
\hline

0.4 ($\rho$)	& 0.4	& 1.5602(2)	& 14-31	& 12/14	& 0.64 \\
0.2 ($\rho$)	& 0.2	& 1.1051(7)	& 12-31	& 14/16	& 0.56 \\
0.1 ($\rho$)	& 0.1	& 0.8620(9)	& 10-31	& 16/18	& 0.57 \\
0.05 ($\rho$)	& 0.05	& 0.7154(17)	& 9-31	& 11/19	& 0.91 \\
0.04 ($\rho$)	& 0.04/0.05	& 0.6853(17)	& 9-28	& 11/16	& 0.83 \\
0.03 ($\rho$)	& 0.03/0.05	& 0.6490(14)	& 8-27	& 19/16	& 0.26 \\
0.02 ($\rho$)	& 0.02/0.05	& 0.6113(19)	& 7-22	& 13/12	& 0.38 \\
0.01 ($\rho$)	& 0.01/0.05	& 0.5730(30)	& 6-18	& 7.4/9	& 0.59 \\
0.04/0.05 ($K^*$)	& 0.04/0.05	& 0.7040(22)	& 9-29	& 15/17	& 0.61 \\
0.03/0.05 ($K^*$)	& 0.03/0.05	& 0.6845(18)	& 9-29	& 21/17	& 0.22 \\
0.02/0.05 ($K^*$)	& 0.02/0.05	& 0.6631(19)	& 8-25	& 15/14	& 0.38 \\
0.01/0.05 ($K^*$)	& 0.01/0.05	& 0.6485(28)	& 8-23	& 4.7/12	& 0.97 \\
0.05 ($\phi$)	& 0.04/0.05	& 0.7198(19)	& 9-29	& 13/17	& 0.77 \\
0.05 ($\phi$)	& 0.03/0.05	& 0.7174(13)	& 9-29	& 19/17	& 0.35 \\
0.05 ($\phi$)	& 0.02/0.05	& 0.7152(11)	& 8-29	& 21/18	& 0.28 \\
0.05 ($\phi$)	& 0.01/0.05	& 0.7194(11)	& 8-28	& 11/17	& 0.84 \\
\end{tabular}
\caption{Vector meson masses.
The format is the same as Table~\protect\ref{PS_MASS_TABLE}.
}
\label{VEC_MASS_TABLE}
\end{center}
\end{table}

\begin{table}
\begin{center}
\begin{tabular}{|llllll|}
$am_{valence}$ & $am_{sea}$ & $am_{SC}$ & range & $\chi^2/D$ & conf. \\
\hline

0.4 ($a_0$)	& $\infty$	& 1.739(4)	& 6-20	& 11/9	& 0.29 \\
0.2 ($a_0$)	& $\infty$	& 1.296(6)	& 6-20	& 5/9	& 0.83 \\
0.1 ($a_0$)	& $\infty$	& 1.063(13)	& 6-20	& 12/9	& 0.24 \\
0.05 ($a_0$)	& $\infty$	& 0.976(28)	& 6-31	& 22/22	& 0.48 \\
0.03 ($a_0$)	& $\infty$	& 0.950(40)	& 5-20	& 9/10	& 0.52 \\
0.02 ($a_0$)	& $\infty$	& 0.848(27)	& 4-20	& 10/11	& 0.51 \\
0.01 ($a_0$)	& $\infty$	& 0.630(40)	& 4-20	& 17/12	& 0.12 \\
\hline

0.02 ($a_0$)	& 0.02	& 0.676(6)	& 4-31	& 25/24	& 0.41 \\
\hline

0.4 ($a_0$)	& 0.4	& 1.750(4)	& 6-20	& 21/9	& 0.013 \\
0.2 ($a_0$)	& 0.2	& 1.297(6)	& 6-20	& 4.8/9	& 0.85 \\
0.1 ($a_0$)	& 0.1	& 1.042(8)	& 6-20	& 18/9	& 0.031 \\
0.05 ($a_0$)	& 0.05	& 0.829(6)	& 5-31	& 63/23	& 1.3e-05 \\
0.04 ($a_0$)	& 0.04/0.05	& 0.808(7)	& 5-20	& 10/10	& 0.41 \\
0.03 ($a_0$)	& 0.03/0.05	& 0.761(12)	& 5-20	& 16/10	& 0.094 \\
0.02 ($a_0$)	& 0.02/0.05	& 0.669(6)	& 4-20	& 31/11	& 0.0011 \\
0.01 ($a_0$)	& 0.01/0.05	& 0.532(8)	& 4-20	& 15/11	& 0.2 \\
\end{tabular}
\caption{$0^{++}$ ($a_0$) meson masses.
The format is the same as Table~\protect\ref{PS_MASS_TABLE}.
}
\label{A0_MASS_TABLE}
\end{center}
\end{table}

\begin{table}
\begin{center}
\begin{tabular}{|llllll|}
$am_{valence}$ & $am_{sea}$ & $am_{PV}$ & range & $\chi^2/D$ & conf. \\
\hline

0.4 ($a_1$)	& $\infty$	& 1.816(7)	& 7-20	& 12/8	& 0.17 \\
0.2 ($a_1$)	& $\infty$	& 1.370(7)	& 6-20	& 4.7/9	& 0.86 \\
0.1 ($a_1$)	& $\infty$	& 1.147(7)	& 5-20	& 3.6/10	& 0.96 \\
0.05 ($a_1$)	& $\infty$	& 1.037(13)	& 5-20	& 7.7/10	& 0.65 \\
0.03 ($a_1$)	& $\infty$	& 0.984(8)	& 4-20	& 15/11	& 0.17 \\
0.02 ($a_1$)	& $\infty$	& 0.950(12)	& 4-18	& 21/9	& 0.013 \\
0.01 ($a_1$)	& $\infty$	& 0.898(19)	& 4-16	& 21/7	& 0.39 \\
\hline

0.02 ($a_1$)	& 0.02	& 0.868(8)	& 4-20	& 11/11	& 0.43 \\
\hline

0.4 ($a_1$)	& 0.4	& 1.826(8)	& 7-20	& 12/8	& 0.17 \\
0.2 ($a_1$)	& 0.2	& 1.388(7)	& 6-20	& 12/9	& 0.2 \\
0.1 ($a_1$)	& 0.1	& 1.149(9)	& 5-20	& 7.3/10	& 0.7 \\
0.05 ($a_1$)	& 0.05	& 1.006(12)	& 5-20	& 28/10	& 0.0018 \\
0.04 ($a_1$)	& 0.04/0.05	& 0.964(10)	& 5-20	& 9.6/10	& 0.47 \\
0.03 ($a_1$)	& 0.03/0.05	& 0.925(6)	& 4-20	& 7.2/11	& 0.78 \\
0.02 ($a_1$)	& 0.02/0.05	& 0.874(6)	& 4-20	& 10/11	& 0.53 \\
0.01 ($a_1$)	& 0.01/0.05	& 8.131(11)	& 4-15	& 2.6/6	& 0.86 \\
\end{tabular}
\caption{$1^{++}$ ($a_1$) meson masses.
The format is the same as Table~\protect\ref{PS_MASS_TABLE}.
}
\label{A1_MASS_TABLE}
\end{center}
\end{table}

\begin{table}
\begin{center}
\begin{tabular}{|llllll|}
$am_{valence}$ & $am_{sea}$ & $am_{PV}$ & range & $\chi^2/D$ & conf. \\
\hline

0.4 ($b_1$)	& $\infty$	& 1.833(9)	& 7-20	& 13/8	& 0.11 \\
0.2 ($b_1$)	& $\infty$	& 1.385(9)	& 6-20	& 8.1/9	& 0.53 \\
0.1 ($b_1$)	& $\infty$	& 1.182(13)	& 5-20	& 12/10	& 0.29 \\
0.05 ($b_1$)	& $\infty$	& 1.110(30)	& 5-20	& 12/10	& 0.3 \\
0.03 ($b_1$)	& $\infty$	& 1.007(15)	& 4-20	& 16/11	& 0.16 \\
0.02 ($b_1$)	& $\infty$	& 0.989(22)	& 4-20	& 13/11	& 0.28 \\
0.01 ($b_1$)	& $\infty$	& 0.990(50)	& 4-16	& 8.9/7	& 0.26 \\
\hline

0.02 ($b_1$)	& 0.02	& 0.903(14)	& 4-20	& 11/11	& 0.43 \\
\hline

0.4 ($b_1$)	& 0.4	& 1.834(12)	& 7-20	& 10/8	& 0.26 \\
0.2 ($b_1$)	& 0.2	& 1.398(10)	& 6-20	& 7/9	& 0.64 \\
0.1 ($b_1$)	& 0.1	& 1.172(12)	& 5-20	& 7.9/10	& 0.64 \\
0.05 ($b_1$)	& 0.05	& 1.047(21)	& 5-20	& 2.5/10	& 0.99 \\
0.04 ($b_1$)	& 0.04/0.05	& 0.996(16)	& 5-20	& 11/10	& 0.39 \\
0.03 ($b_1$)	& 0.03/0.05	& 0.957(9)	& 4-20	& 9.5/11	& 0.58 \\
0.02 ($b_1$)	& 0.02/0.05	& 0.915(13)	& 4-20	& 13/11	& 0.3 \\
0.01 ($b_1$)	& 0.01/0.05	& 0.856(19)	& 4-18	& 8.7/9	& 0.47 \\
\end{tabular}
\caption{$1^{+-}$ ($b_1$) meson masses.
The format is the same as Table~\protect\ref{PS_MASS_TABLE}.
}
\label{B1_MASS_TABLE}
\end{center}
\end{table}

\begin{table}
\begin{center}
\begin{tabular}{|llllll|}
$am_{valence}$ & $am_{sea}$ & $am_{B}$ & range & $\chi^2/D$ & conf. \\
\hline

0.4 (N)	& $\infty$	& 2.3933(15)	& 18-31	& 21/10	& 0.021 \\
0.2 (N)	& $\infty$	& 1.6847(23)	& 16-31	& 19/12	& 0.084 \\
0.1 (N)	& $\infty$	& 1.2770(50)	& 14-30	& 15/13	& 0.33 \\
0.05 (N)	& $\infty$	& 1.0470(70)	& 12-21	& 2.9/6	& 0.82 \\
0.03 (N)	& $\infty$	& 0.9400(50)	& 9-19	& 8.1/7	& 0.32 \\
0.02 (N)	& $\infty$	& 0.8820(70)	& 8-17	& 5.3/6	& 0.5 \\
0.01 (N)	& $\infty$	& 0.8100(90)	& 7-14	& 6.8/4	& 0.15 \\
0.03/0.05 ($\Lambda$)	& $\infty$	& 0.9730(40)	& 8-19	& 5.7/8	& 0.68 \\
0.02/0.05 ($\Lambda$)	& $\infty$	& 0.9310(40)	& 7-19	& 7.1/9	& 0.63 \\
0.01/0.05 ($\Lambda$)	& $\infty$	& 0.8900(60)	& 7-15	& 11/5	& 0.05 \\
0.03/0.05 ($\Xi$)	& $\infty$	& 1.0090(30)	& 8-20	& 6.4/9	& 0.7 \\
0.02/0.05 ($\Xi$)	& $\infty$	& 0.9867(28)	& 7-20	& 6.6/10	& 0.76 \\
0.01/0.05 ($\Xi$)	& $\infty$	& 0.9660(40)	& 7-18	& 9.4/8	& 0.31 \\
\hline

0.02 (N)	& 0.02	& 0.8450(40)	& 8-20	& 16/9	& 0.074 \\
0.02/0.05 ($\Lambda$)	& 0.02	& 0.9100(30)	& 7-21	& 16/11	& 0.14 \\
0.02/0.05 ($\Xi$)	& 0.02	& 0.9752(23)	& 7-22	& 17/12	& 0.15 \\
\hline

0.4 (N)	& 0.4	& 2.4213(19)	& 18-31	& 4.6/10	& 0.92 \\
0.2 (N)	& 0.2	& 1.7075(22)	& 16-31	& 17/12	& 0.15 \\
0.1 (N)	& 0.1	& 1.3110(40)	& 14-28	& 17/11	& 0.12 \\
0.05 (N)	& 0.05	& 1.0570(50)	& 12-25	& 5/10	& 0.89 \\
0.04 (N)	& 0.04/0.05	& 1.0030(30)	& 10-22	& 17/9	& 0.044 \\
0.03 (N)	& 0.03/0.05	& 0.9300(27)	& 9-22	& 7.5/10	& 0.68 \\
0.02 (N)	& 0.02/0.05	& 0.8540(30)	& 8-20	& 6.9/9	& 0.65 \\
0.01 (N)	& 0.01/0.05	& 0.7790(60)	& 7-16	& 1.1/6	& 0.98 \\
0.04/0.05 ($\Lambda$)	& 0.04/0.05	& 1.0240(30)	& 9-22	& 9/10	& 0.53 \\
0.03/0.05 ($\Lambda$)	& 0.03/0.05	& 0.9784(23)	& 8-21	& 7/10	& 0.72 \\
0.02/0.05 ($\Lambda$)	& 0.02/0.05	& 0.9312(27)	& 7-20	& 9.9/10	& 0.45 \\
0.01/0.05 ($\Lambda$)	& 0.01/0.05	& 0.8850(50)	& 7-18	& 4.2/8	& 0.84 \\
0.04/0.05 ($\Xi$)	& 0.04/0.05	& 1.0440(30)	& 9-22	& 8.3/10	& 0.6 \\
0.03/0.05 ($\Xi$)	& 0.03/0.05	& 1.0214(20)	& 8-22	& 8.3/11	& 0.68 \\
0.02/0.05 ($\Xi$)	& 0.02/0.05	& 0.9989(20)	& 7-21	& 9.4/11	& 0.58 \\
0.01/0.05 ($\Xi$)	& 0.01/0.05	& 0.9798(27)	& 7-20	& 6.5/10	& 0.77 \\
\end{tabular}
\caption{Octet baryon masses.
The format is the same as Table~\protect\ref{PS_MASS_TABLE}.
}
\label{BARYON_MASS_TABLE}
\end{center}
\end{table}

\begin{table}
\begin{center}
\begin{tabular}{|llllll|}
$am_{valence}$ & $am_{sea}$ & $am_{\Delta}$ & range & $\chi^2/D$ & conf. \\
\hline

0.4 ($\Delta$)	& $\infty$	& 2.424(1)	& 10-25	& 39/12	& 0.0001 \\
0.2 ($\Delta$)	& $\infty$	& 1.737(2)	& 9-25	& 30/13	& 0.0053 \\
0.1 ($\Delta$)	& $\infty$	& 1.373(6)	& 8-17	& 6.9/6	& 0.33 \\
0.05 ($\Delta$)	& $\infty$	& 1.184(12)	& 7-13	& 0.37/3	& 0.95 \\
0.03 ($\Delta$)	& $\infty$	& 1.092(12)	& 5-11	& 3.5/3	& 0.32 \\
0.02 ($\Delta$)	& $\infty$	& 1.058(15)	& 4-9	& 0.85/2	& 0.65 \\
0.01 ($\Delta$)	& $\infty$	& 1.008(17)	& 3-7	& 1.7/1	& 0.19 \\
\hline

0.02 ($\Delta$)	& 0.02	& 0.991(18)	& 4-10	& 1.7/3	& 0.65 \\
\hline

0.4 ($\Delta$)	& 0.4	& 2.450(6)	& 18-25	& 6.2/4	& 0.19 \\
0.2 ($\Delta$)	& 0.2	& 1.753(2)	& 9-23	& 12/11	& 0.33 \\
0.1 ($\Delta$)	& 0.1	& 1.396(6)	& 8-17	& 11/6	& 0.08 \\
0.05 ($\Delta$)	& 0.05	& 1.155(8)	& 7-15	& 6.4/5	& 0.27 \\
0.04 ($\Delta$)	& 0.04/0.05	& 1.102(9)	& 6-14	& 8.9/5	& 0.11 \\
0.03 ($\Delta$)	& 0.03/0.05	& 1.059(6)	& 5-13	& 5.8/5	& 0.32 \\
0.02 ($\Delta$)	& 0.02/0.05	& 0.989(7)	& 4-11	& 10/4	& 0.04 \\
0.01 ($\Delta$)	& 0.01/0.05	& 0.921(12)	& 3-9	& 8.8/3	& 0.032 \\
\end{tabular}
\caption{Decuplet baryon masses.
The format is the same as Table~\protect\ref{PS_MASS_TABLE}.
}
\label{DELTA_MASS_TABLE}
\end{center}
\end{table}


\begin{table}
\begin{center}
\begin{tabular}{|lllll}
\hline
$a m_{valence}$ & $am_{sea}$ & $c_\pi(0,0,1)$ & $c_\pi(0,1,1)$ & $c_\rho(0,0,1)$ \\
\hline

0.10	& $\infty$	& 0.981(6) & 0.971(5) & 0.984(49) \\
0.02	& $\infty$	& 1.001(10) & 0.973(21) & 0.900(82) \\
0.01	& $\infty$	& 0.996(14) & 0.981(59) & 1.021(122) \\
\hline

0.02    & 0.02  & 0.997(11) & 0.986(16) & 1.013(62) \\
\hline

0.4	& 0.4	& 0.966(19) & na & 0.949(16) \\
0.2	& 0.2	& 0.958(8) & 0.952(4) & 0.921(30) \\
0.1	& 0.1	& 0.982(5) & 0.976(5) & 0.954(32) \\
0.05	& 0.05	& 0.996(6) & 0.988(9) & 0.952(44) \\
0.04	& 0.04/0.05	& 1.000(8) & 0.972(9) & 0.926(51) \\
0.03	& 0.03/0.05	& 0.988(5) & 0.978(9) & 0.903(28) \\
0.02	& 0.02/0.05	& 1.001(8) & 0.993(16) & 0.982(60) \\
0.01	& 0.01/0.05	& 0.995(17) & 0.986(30) & 0.967(63) \\
\end{tabular}
\caption{``Speed of light'' for the pion and rho at $a \approx 0.13$ fm.
The three columns are for the Goldstone pion with momenta
$\frac{2\pi}{L}\LP 0,0,1 \RP$ and $\frac{2\pi}{L}\LP 0,1,1 \RP$ and the $\gamma_z \otimes
\gamma_z$ rho with momentum $\frac{2\pi}{L}\LP 0,0,1 \RP$.  
}
\label{C_TABLE}
\end{center}
\end{table}

\end{document}